\documentclass[useAMS,usenatbib]{mn2e}
\usepackage{amsmath,amssymb,wasysym}
\usepackage{delarray}
\usepackage{graphicx}
\usepackage{tabularx}
\newcommand{\intd}{{\rm{d}}}  % integrand d (dstraight)

\newcommand{\Rv}{{R_{\rm vir}}}
\newcommand{\Mtr}{{M_{\rm tr}}}
\newcommand{\Mdot}{{\dot{ M}}}
\newcommand{\mdot}{{\dot{ m}}}

\newcommand{\Ms}{{M^{\bigstar}}}
\newcommand{\Mstellar}{\Ms}
\newcommand{\Mq}{{M_{\rm quench}}}
\newcommand{\Mqb}{{{M_{\rm quench B06}}}}

\newcommand{\Msun}{\ensuremath{\mathrm{M}_{\odot}}}
\newcommand{\MDM}{{M_{\rm DM}}}
\newcommand{\Mps}{{ M_{\ast}}}
\newcommand{\Mst}{{M_{\rm stream}}}
\newcommand{\MstDB}{{M_{\rm streamDB06}}}

\newcommand{\Msh}{{M_{\rm shock}}}

\newcommand{\Zsun}{Z_{\odot}}

\newcommand{\Zhot}{{{Z_{\rm hot}}}}
\newcommand{\fhot}{{{f_{\rm hot}}}}
\newcommand{\fcold}{{{f_{\rm cold}}}}

\newcommand{\Sec}[1]{Sect.~\ref{s:#1}}
\newcommand{\Fig}[1]{Fig.~\ref{f:#1}}

\newcommand{\M}[1]{\mathbf{#1}}         % matrix notation
\newcommand{\pdrv}[2]{\frac{\partial #1}{\partial #2}} % partial deriv
\newcommand{\pdrvd}[3]{\frac{\partial^{2} #1}{{\partial #2}{\partial #3}}} % double partial deriv
\newcommand{\nicefrac}[2]{\leavevmode\kern.1em
            \raise.5ex\hbox{\the\scriptfont0 #1}\kern-.1em
      /\kern-.15em\lower.25ex\hbox{\the\scriptfont0 #2}}

\begin{document}

%\title[The regimes of gas accretion on galaxies in cosmological simulations]{The regimes of gas accretion on galaxies in cosmological simulations}

\title{Bimodal gas accretion in the Horizon-MareNostrum galaxy formation simulation}
%{Bimodal gas accretion in cosmological simulations}

\author[P.\ Ocvirk, C.\ Pichon, R. Teyssier]{P. Ocvirk$^{{1,2}}$, C.\ Pichon$^{{3,2}}$ \& R.\ Teyssier$^{{2}}$\\
$^1$Astrophysikalisches Institut Potsdam, An der Sternwarte 16, D-14482 Potsdam, Germany. \\
$^2$  Institut de Recherches sur les lois Fondamentales de l'Univers, DSM,  l'Orme des Merisiers, 91198 Gif-sur-Yvette, France. \\
$^3$Institut d'Astrophysique de Paris (UMR7095) et UPMC, 98 bis boulevard Arago, 75014 Paris, France. \\
%$^3$Numerical Investigations in Cosmology (N.I.C.), CNRS, France.\\
}

%\begin{document}

\date{Typeset \today; Received / Accepted}

%\titlerunning{Stellar content from high resolution galaxy spectra}
\pagerange{\pageref{firstpage}--\pageref{lastpage}} \pubyear{2008}

\maketitle
\label{firstpage}

%\tableofcontents

%\onecolumn

%\begin{document}

\begin{abstract}
The physics of diffuse gas accretion and the
properties of the cold and  hot modes of accretion onto proto-galaxies
between  $z=2$  and  $z=5.4$ is investigated using 
the  large cosmological simulation performed with
the  RAMSES  code  on  the  MareNostrum  supercomputing  facility.   
 Galactic  winds,  chemical  enrichment,  UV
background heating and radiative  cooling are  taken into account
in    this   very    high   resolution    simulation.     Using   {\it
accretion--weighted  temperature  histograms},  we have  perfomed  two
different  measurements  of the  thermal  state  of  the gas  accreted
towards  the   central  galaxy.   \\
 The  first  measurement,   performed  using
accretion--weighted  histograms on a spherical surface of  radius  0.2  $\Rv$
centred  on {the densest gas structure near the halo centre
  of  mass}, is  a good  indicator of  the presence of an  accretion shock in the vicinity  of the galactic disc.
We  define  the hot  shock  mass, $\Msh$,  as  the  typical halo  mass
separating cold dominated from hot dominated accretion  in the vicinity
of  the  galaxy.  \\
The  second  measurement  is  performed by  radially
averaging histograms between  $0.2 \Rv$ and $\Rv$, in  order to detect radially
extended structures  such as gas filaments:  this is a  good proxy for
detecting cold  streams feeding the central galaxy.   We define $\Mst$
as  the transition  mass separating  cold dominated  from  hot dominated
accretion in  the outer  halo, marking the  disappearance  of  these cold
streams. \\
  We find  a  hot shock  transition  mass of  $\Msh=10^{11.6} \,
\Msun$ (dark matter),  with no significant  evolution with  redshift. Conversely,  we find  that $\Mst$ increases
sharply with $z$.  Our measurements are in agreement with the analytical predictions of \citet{BD03} and
\citet{DB06}, {\em if we correct their model by assuming low metallicity ($\le 10^{-3} \, \Zsun$) for the filaments, correspondingly to our measurements.}
%{\it  if one takes  properly into account the  effect of
%metals  when computing  radiative cooling  rates}.\\
Metal  enrichment of  the intergalactic medium is therefore  a key
ingredient in  determining the transition mass from cold to hot  dominated diffuse  gas  accretion. \\
We find that the diffuse cold gas supply at the inner halo stops at $z=2$ for objects with stellar masses of about $10^{11.1} \Msun$, which is close to the quenching mass determined observationally by \cite{bundy06}. However, its evolution with $z$ is not well constrained, making it difficult to rule out or confirm the need for an additional feedback process such as AGN. 
\end{abstract}

\begin{keywords}
  methods:  Numerical  simulations,  N-body, hydrodynamical,  adaptive
  mesh refinement, galaxies: formation
\end{keywords}
%
%________________________________________________________________

\setcounter{page}{1}

%\tableofcontents

%\vfill

%%%%%%%%%%%%%%%%%%%%%%%%%%%%%%%%%%%%%%%%%
\section{Introduction}
%%%%%%%%%%%%%%%%%%%%%%%%%%%%%%%%%%%%%%%%%

It is currently accepted that the $\Lambda {\rm {CDM}}$ theory provides
a framework with which a large  number of observed galaxy properties can be
interpreted. This framework is referred to as the ``hierarchical scenario of galaxy
formation''.   Most  importantly,  this framework   explains  why  many  of  these
properties (physical sizes, black  hole mass, bulge mass...) are found
to correlate simply  with galaxy mass \citep{KH00}. Amidst this
apparently  simple  scaling  of   galaxy  properties  with  mass,  the
discovery of  a bimodality in  the colour distribution of  Sloan Digital Sky Survey (SDSS)
galaxies  \citep{kauffmann03} stood  unexpected and  at odds  with the
predictions of  hierarchical galaxy formation.   Galaxy bimodality can
indeed appear to be anti-hierarchical, as  can be grasped from the following
simple argument. Under the  assumption (common in early semi-anlytical
models, hereinafter  SAMs) that the star formation  rate (hereafter SFR)
is  proportional   to  the  gas  accretion  rate,   the  latter  being
proportional to the halo mass to some power \citep{vandenbosch02}, one
expects that, at all times, objects with the highest SFR should be the
largest  galaxies.   In this  framework,  massive elliptical  galaxies
would  still  be blue  and  forming stars  at  $z=0$. The fact  that
observed elliptical  galaxies do not obey  this fundamental prediction
of  the   hierarchical  scenario  is  the  origin   of  the  so-called
``anti-hierarchical'' behaviour of massive red galaxies \citep{rasera06}. This observation is further supported by the analysis of spectroscopic data, using star formation history reconstruction methods \citep{moped01,panter1,CF04-1,stecmap,steckmap}.
Since these giant galaxies are in the form  of apparently ``dead''  (i.e.  no ongoing
star formation)  red elliptical  galaxies, the quest  has been  ongoing for
several years  to find the origin  of this halt in  the star formation
process  (also  refered  to   as  ``star  formation  quenching'').   

A substantial  part of astrophysical  research nowadays is devoted to
searching for new physical mechanisms able to prevent cold gas accreted at
$\Rv$ from  falling into the galactic disc,  condensing into molecular
clouds and forming  stars. The heating of the infalling gas from virialization and feedback from supernovae and hot stars has been considered as a serious candidate for more than a decade but seems insufficient to explain the drop in star formation of massive systems in recent times \citep{rasera06}, leading the authors to suggest a superwind phase for the high mass end of the galaxy population.

Active Galactic Nuclei (AGN) feedback has been  proposed by several
authors  \citep{bower06,hopkins07} as the origin of this quenching, and  has the  additional
desirable propery of preventing cooling  flows in the core of the most
massive  cluster galaxies \citep{delucia06,cattaneo07}.   However, AGN
physics are still poorly understood, both theoretically (because of the
intrinsic complexity of  relativistic magnetohydrodynamic flows around
black holes, see for instance \cite{progareview07}), and observationally (because of the small physical extent of the
region of  interest). Moreover, the mechanism involved in transferring the energy from the black hole accretion flow to the surroundings remains elusive. Jets have been proposed, and have the advantage of being supported by observations, but shock waves arising from the interaction of the jet with the interstellar medium would tend to push away the hot gas while leaving surrounding clumps of cold gas rather unchanged \citep{slyz05}. However, this would depend on the position of the clumps with respect to the jet origin and the violence of the shock, and for instance, with a Mach number $\approx 10$ and density contrast $\approx 10$, \cite{nakamura06} do indeed predict cloud destruction.

 A jet-driven turbulence is another alternative, and to assess its relevance one has to repeat experiments such as those of  \cite{banerjee07} and \cite{cattaneo07} at the galactic scale. The balance between the mechanical power output of bubble-carving jets (estimated from radio luminosity at 1.4 GHz) and the radiative losses of the hot gas halo of galaxy clusters has been proposed as a signature of the global control of gas cooling by interaction with a central black hole jet \citep{best06}. However, a necessary requirement for this scenario to work for galaxies is that the energy available in the jet remains in the galaxy, while observations show that radio sources have linear sizes significantly larger than their host galaxies. It is thus not clear how jets can prevent star formation, and it has also been argued that they might actually {\em enhance} star formation \citep{silk05}. More generally, the causal relation between AGNs and star formation is unclear: does the starburst trigger AGN activity and a subsequent quenching or does the AGN activity trigger the starburst? As a matter of fact, strong AGN activity is seen in galaxies with intense star formation activity \citep{wild07,cidfernandes01}, demonstrating that AGN and star formation co-exist, although this could be just a short-lived phase \citep{kauffmann03,schawinski07,ciotti07}. Finally, purely radiative feedback from the accretion disc around the black hole might be an alternative \citep{fabian06,fabian08}, but the mechanical coupling through which the dust phase being blown away drags the cold gas along is uncertain.
Hence, it is worth  investigating possible
quenching  mechanisms other than AGN feedback. 
In  this  respect, the  detailed analysis  of
diffuse  gas  accretion  around  star  forming galaxies  is  of  great
interest  because  it can  provide  a  form  of self-regulation.   The
seminal  paper  of   \cite{BD03}  (hereinafter  BD03)  investigates  the
stability of  hot accretion shocks around disc  galaxies, showing that
such  shocks can  exist  only  for haloes  more  massive than  $\approx
10^{11.5} \Msun$.   In an ideal  spherical flow, this hot  shock would
prevent cold gas from reaching the disc (or at least slow it down) and
thus is likely to  affect   star  formation.   \cite{DB06}   (hereinafter  DB06)
extended this approach  to the study of the  stability of cold streams
(``filaments'') within  the shock--heated  halo gas. They  showed that
the observed transition mass from blue to red galaxies at $z \simeq 0$
could  be matched to  the critical  mass at  which a  stable accretion
shock can  exist and that stable filaments would disappear around $z=1.5$. These  findings were also  driven and further confirmed by
numerical simulations  of high redshift  galaxy formation based  on smoothed particle hydrodynamics (SPH), as in \cite{keres05}. DB06 actually presented the rise of stable hot shocks not as the origin of the quenching but only as a necessary condition for an efficient AGN feedback.

 However, it is too early to discard the existence of stable hot shocks or the destruction of the gas filaments as the origin of the galaxy bimodality. Indeed, no numerical  study has  considered  the  influence  of  chemical
enrichment, which was shown in DB03  and DB06 to have a crucial impact
on  shock  stability,  since  metallicity,  along  with  gas  density,
determines the  cooling rate \citep{SD93}. More recently, \cite{cattaneob07} checked the good behaviour of SAMs with respect to hot and cold accretion modes by comparing GalICS \citep{hatton03} results to the \cite{keres05} simulation, which does not take into account chemical enrichment. 

In this  paper, we address
these  very issues  using the  results  of a  very large  cosmological
simulation performed on the MareNostrum supercomputer at the Barcelona
Supercomputer  Centre.   It  was  performed  within  the Horizon
collaboration  ({\tt http://www.projet-horizon.fr})  using  the RAMSES  code
\citep{teyssier02},  which now includes  a detailed treatment of metal--dependent
gas cooling, UV heating, star formation, supernovae feedback and metal
enrichment. 

The  simulation parameters  ($L_{\rm box}=50$  $h^{-1}$Mpc, $1024^3$  dark
matter particles and a spatial  resolution of about 1 $h^{-1}$kpc) are
optimal  to capture  the most  important properties  of  gas accretion
around typical Milky Way--like galaxies.  The large box size allows us
to  have a large  (more than  100) sample  of $L_{\star}$  galaxies at
$z\ge2$ and above,  with a  strong statistical  significance.  For
this  mass  scale,  the  main  transitions  in  the  accretion  regime
(appearance of  hot accretion  shocks, disappearance  of  cold filaments)
effectively  takes place  between $2\le z \le 6$.   Moreover,  the adaptive mesh refinement (AMR) method   adopted  in  RAMSES  allows  us  to
investigate  the flow  in an  Eulerian framework,  in contrast  to the
Lagrangian  approach  adopted  by  \cite{keres05}.   Finally,  it  was
recently  observed  that  the  apparent bimodality  mass  scale  might
increase  with   redshift  \citep{juneau05,  bundy06,   hopkins07},  an
intriguing behaviour that has yet to be checked against models.

The  outline  of  this  paper  is  as  follows:  first  we  describe  in Sec. 2 our
methodology,   in  terms   of  numerical   techniques   and  statistical
measurements.   We specifically introduce a new  estimator to  analyse the
thermodynamical     properties    of     accretion,     namely    {\it
accretion--weighted  histograms}.  We  then present  in Sec. 3 our  main results
concerning the physical properties  of the accreted gas.  Our findings
are then  discussed in the framework of  earlier theoretical modelling in Sec. 4, and recent observations in Sec. 5.

%%%%%%%%%%%%%%%%%%%%%%%%%%%%%%%%%%%%%%%%%
\section{Methodology}
%%%%%%%%%%%%%%%%%%%%%%%%%%%%%%%%%%%%%%%%%

\label{s:method}

In this section, we first describe the MareNostrum simulation: a
cosmological  N  body and  hydrodynamics  simulation of  unprecedented
scale with most of the physical processes involved in galaxy formation
theory. We  then describe our  dark matter halo catalogue and its corresponding properties, 
 present our  statistical  tool --
accretion--weighted   temperature  histograms,  and   discuss our criterion for separating diffuse gas accretion from satellite merging.

\subsection{The Horizon-MareNostrum simulation}
\label{s:MN}

We have  performed a  cosmological simulation of  unprecedented scale,
using  2048 processors of  the MareNostrum  computer installed  at the
Barcelona Supercomputing Centre in Spain. We have used intensively the
AMR  code RAMSES \citep{teyssier02}  for 4  weeks dispatched  over one
full year.   This effort is part  of a consortium  between the Horizon
project in  France ({\tt http://www.projet-horizon.fr}) and  the MareNostrum
galaxy          formation           project          in          Spain
({\tt http://astro.ft.uam.es/$\sim$marenostrum}). This simulation should therefore be named the Horizon-MareNostrum simulation to avoid confusion with the GADGET-2 MareNostrum simulation \cite{gottloeber07}. In the rest of the paper, we refer exclusively to the Horizon-MareNostrum simulation unless explicitly stated otherwise.

The main asset  of this
project relies  on using  a  quasi exhaustive number of  physical ingredients
that are  part of the current  theory of galaxy formation,  and at the
same time covering a large enough  volume to provide a fair sample of the
universe, especially  at redshifts above one.  
%The  physical processes we
% have included in our simulation  are described now in more details. 
Specifically, we
have considered metal-dependent cooling and UV heating using the Hardt
and Madau  background model.  We  have incorporated a simple  model of
supernovae  feedback  and metal  enrichment  using the  implementation
described  in  \citet{Dubois08}.  For  high--density  regions, we  have
considered  a  polytropic equation  of  state  with a 5/3 index to  model the  complex,
multi-phase  and  turbulent   structure  of  the  ISM  \citep{Yepes97,
Springel03} in a simplified form (see \citet{Schaye07, Dubois08}): the
ISM is defined as gas with a density greater than $n_0 \simeq 0.1$ H/${\rm cm^3}$. Star
formation has  also been  included, for ISM  gas only  ($n_{\rm H}>n_0$), by
spawning star  particles at  a rate consistent  with the  Kennicutt law
derived  from local observations  of star  forming galaxies.   Technically, 
we have  $\dot \rho_* = \rho_{gas}/t_*$ where $t_*
= t_0  (n_{\rm H}/n_0) ^{-1/2}$  and $t_0=8$ Gyr.  Recast in units  of the
local free-fall time, this  corresponds to a star formation efficiency
of 5\%. The simulation was started  with a base grid of $1024^3$ cells
and  the  same number  of  dark matter  particles,  and  the grid  was
progressively  refined,  on a  cell--by--cell  basis,  when the  local
number of particles exceeded 10.   A similar citerion was used for the
gas,  implementing  what   is  called  a  Quasi-Lagrangian  refinement
strategy.  Five additional levels of refinement were  considered, but the
maximum level of refinement was adjusted so that the minimum cell size
in {\it physical units} never exceeded one kpc. In this way, our spatial
resolution is  consistent with the angular resolution  used to derived
the Kennicutt law from observations. On  the other hand, we are not in
a position to resolve the scale height of thin cold discs so the
detailed galactic dynamics are likely to be be affected by resolution effects.

The   simulation   was   ran   for  a   $\Lambda$CDM   universe   with
$\Omega_M=0.3$,  $\Omega_{\Lambda}=0.7$, $\Omega_B=0.045$,  $H_0=70$ km/s/Mpc,
$\sigma_8=0.9$ in  a periodic box  of 50 $h^{-1}$Mpc. Our  dark matter
particle  mass ($m_{\rm part.}  \simeq  8\times 10^6$  $M_{\odot}$), our  spatial
resolution (1 kpc {\it physical})  and our box size make this simulation
ideally suited to  study the formation of galaxies  within dark matter
haloes, from dwarf-- to Milky Way--sized objects at high redshift. For
large galaxies,  we can nicely resolve  the radial extent  of the disc, but 
not its  vertical extent, while  for small galaxies, we  can resolve
the gravitational contraction of the cooling gas, but barely the final
disc. 

A common goal of the Horizon collaboration and the MareNostrum galaxy formation simulation project is to investigate the relative accuracy of the current SPH and AMR codes, through direct comparison of the results of a large galaxy formation simulation.
To allow for this, the Horizon-MareNostrum simulation and the GADGET-2 MareNostrum simulation use the same initial conditions, which  are described in \cite{gottloeber07} and  \cite{prunet08}.
 The simulation was stopped  at redhift $z \simeq 1.5$ because the  allocated time ran out. The total number of galaxies at the
end of the simulation was larger than  $2 \times 10^5$, the total number of star particles was more than $10^8$, and the total number of AMR cells was larger than $5\times 10^9$.

\subsection{Virial spheres at rest}
\label{s:Virialdef}

In  order to  analyse  the physical  properties  around high  redshift
galaxies, we have built  from our simulation data a Friend--Of--Friend
(FOF)  halo catalogue  \citep{efstathiou88}. For  each snapshot  and for
each halo,  we compute its  mass and its  centre of mass.   The Virial
radius is defined here as  $R_{200b}$, the radius at which the average
mass density in  the halo is 200 times  the background matter density.
We then  define the  Bright Central  Galaxy (BCG) of  the halo  as the
highest gas  density peak in a  sphere centred on the  halo centre of
mass  and  of radius  $0.5  \Rv$.  This  position  will  serve as  our
reference  point  for computing  radial  accretion  rates.  We checked  a
posteriori  that this position  also corresponds  to the  most massive
substucture in the halo.  We  subtract from the gas velocity the mass-averaged
velocity  of the  gas inside  the Virial  radius in  order to  put the
system at rest.  The basic physical  properties of the gas are then
mapped onto concentric shells centred on the BCG position. 
%\pier{une jolie
%  Fig. aitoff temperature et mdot, montrant comment on accrete plus la ou
%  c'est froid....}

%\subsection{A new tool: the ${\dot{\bf M}}$-weighted probability distribution function}
\subsection{A new tool: accretion--weighted histograms}

It is quite common in cosmological simulations including gas physics to
analyse the thermodynamical state  of baryons using so--called ``phase
space  diagrams''\footnote{as in different physical phases, {\em not} position-velocity phase space} \citep{cen93,katz96,rasera06},  for which  the total
mass fraction in a given gas density and temperature range is given as
2D histograms.   This sort  of diagram yields  only a static  view and
does not include any reference to mass or energy fluxes.  Instead, we
propose  to gain insight  into the  accretion regimes  of cosmological
haloes by using a  new tool: accretion--weighted phase space diagrams.
We  still   use  temperature  and   density  probability  distribution
functions,  but   we  weight  the  contribution  to  each
temperature and density bin by  the local accretion rate. In this way,
static  regions  will be  discarded  from  the  analysis, while  large
radial velocity  regions will  dominate the  signal. Since  the  accretion is
towards  the BCG in the  halo  centre, we  define
concentric shells where the temperature, density, velocity and metallicity
are computed by smoothing the underlying 3D fields with a window function
of scale $R$:
\begin{equation}
T_R(r,\theta,\phi) = \int T(\M{x}')W_R(\M{x}-\M{x}') \intd^3\M{x}'\,.
\end{equation}
We  then  sample this  3D  field on  various  spherical surfaces  of radius  $0.2
R_{\rm vir}<r<R_{\rm vir}$, with an angular resolution $\Delta \theta \simeq R
/ r$.   In the current  analysis, we fix  $R \simeq 2$kpc,  twice our
spatial  resolution, so  that for  our largest  haloes, the  sphere was
sampled with $400\times 400$ pixels. We will thus drop the subscript $_R$ for the rest of the paper. We obtained for each halo angular
maps of  each smoothed gas variables  (density $\rho$, velocity $v$, temperature ${\rm T}$,
metallicity $Z$).
We then define the local accretion rate $\mdot$ as (see
e.g. \cite{aubert}, \cite{pichon07})
\begin{equation}
\mdot (r,\Omega)= 
%\frac{\intd \Mdot}{\intd \Omega} 
\pdrv{\Mdot}{ \Omega} 
= 
\rho \, \M{v} \cdot \M{n} \, r^2 \, , \label{eq:defMDR}
\end{equation}
where the solid  angle is defined by $\intd  \Omega= \sin \theta \intd
\theta \intd \phi$ in the direction $\Omega=(\theta,\phi)$, and $\M{n}$ is the vector normal to the sphere and of norm 1. The total accretion rate  across the sphere is
recovered using  $\Mdot =  \int \mdot \, \intd\Omega$. 
%\pier{eventuellement
%  on pourrait caser la Fig. du Mdot vs Mass ici mais faudrait prendre celle
%  qui est clumpy$+$diffuse} 
For most of this  paper, we will omit
the density  variable, $\rho$,  to focus on  the thermal  and chemical
properties of the  accretion flow, described here by  $T$ and $Z$.
 At a  given radius,
$r$, we  can marginalize Eq. (\ref{eq:defMDR}) over all cells which have 
a given temperature, $T$,  and obtain  the accretion
rate per unit temperature as
\begin{equation}
\mdot(r,T) = 
\int \delta_{\rm D}(T-T(\Omega)) \, \mdot \, \intd \Omega
=\pdrv{\Mdot}{ T}\, , \label{eq:defMDRT}
\end{equation}
where $\delta_{\rm D}$ is the Dirac function. The total  accretion rate across the sphere  is recovered  using $\Mdot =  \int \mdot \, \intd T$.
Similarly, while marginalizing over all angles which have   a given temperature, $T$, 
and a given metallicity, $Z$,  we may introduce our     main    tool,        the    accretion-weighted
temperature--metallicity two-dimensional probability density function
(hereinafter PDF)
\begin{equation}
\mdot(r,T,Z)\!\! = \!\!
\int \!\! \delta_{\rm D}(T-T(\Omega)) \delta_{\rm D}(Z-Z(\Omega))  \mdot\intd \Omega = \pdrvd{\Mdot}{T}{Z}\,.
\label{eq:defMDRZ}
\end{equation}
%From now on, we drop the subscript $R$, 
%since in the present analysis it is fixed to the physical
%resolution of the simulation.

\subsection{Diffuse accretion versus clumpy satellites}
\label{s:trunc}
%\pier{ Reformulated, please reread: 
In this  paper, we are  interested in characterizing the  accretion of
diffuse  intergalactic  gas  rather   than  the  accretion  of  galaxy
satellites.   We  therefore  need  to  separate  the  contribution  of
infalling gaseous discs from the smooth accretion through filaments or
other diffuse  components.  In our model, the  star--forming dense ISM
is defined as $n_{\rm H} \ge$ $0.1 \rm{H/cm^3}$. We remove from our spherical
analysis all pixels whose density exceeds this theshold.  Although this
truncation may seem brutal, infalling satellites  are
frequently embedded in diffuse filaments. As such, the filament is the
natural surrounding  of the infalling satellite, and deciding where the
boundary lies is difficult yet critical. In this context, a density criterion is still the
most straightforward separation method. In the future, a possible alternative could be to use a
segmentation algorithm to separate the filament from the satellites on  topological grounds, 
in the spirit of the skeleton reconstructions of \cite{sousbie08}.
%}
%\pier{put fig. 7 here? or remove it totally?}

\subsection{Entire halo versus galaxy vicinity}

We use  two distinct estimators, $\mdot$ and $\langle \mdot \rangle$,
  to characterize the diffuse gas accretion
around the  central galaxy. 
\\
The first  one, $\mdot$ is directly  related to the
gas   properties  close  to   the  galactic   disc:  we   compute  the
accretion-weighted $T-Z$ histogram, Eq. (\ref{eq:defMDRZ}),  at radius $r=0.2$ $\Rv$, which turned
out  to be close  enough to, but  not intersecting,  the neutral  HI disc.
This  region defines the  disc vicinity,  for which  we would  like to
analyse the thermal  properties of the accreted gas.  We easily detect
the presence of an accretion shock  upwind of the disc if the accreted
gas  is predominantly  in  a hot  phase.  On the  other  hand, if  the
accreted  gas is  predominantly  in a  cold  phase, it  means that  no
accretion shock is present above $0.2R_{\rm vir}$. 
\\
The second estimator, $\langle \mdot \rangle$, is
based  on  averaging  the  accretion-weighted histograms  measured  at
different radii between $0.2 \Rv$ and $\Rv$.1
\begin{equation}
\langle \mdot \rangle(T,Z) = \int_{0.2R_{\rm vir}} ^{R_{\rm vir}} \mdot(r,T,Z)\frac{{\rm d} r}{\Delta r}\,, \label{eq:defRMD}
\end{equation}
where $\Delta r\equiv 0.8$ $\Rv$ and $\mdot(r,T,Z)$ is given by  Eq. (\ref{eq:defMDRZ}).
Coadding different histograms at  different radii increases the weight
of  coherent  radial structures  such  as  filaments  or  cold  streams  that
eventually extend  out to  (or beyond) the  Virial radius.   Indeed, 
if the velocity flow is chemo-thermodynamically similar between two  neighbouring shells, the corresponding histograms will add 
up consistently.
 Note that
under  the  assumption of  a  steady  spherically  symmetric flow,  the
accretion rate  does not depend on  radius.  As  we will  see later  on, even in  the presence  of an
accretion shock,  cold streams  can persist in  the halo  and directly
feed the central galaxy with fresh cold gas. This phenomenon will appear
in  our histograms as a dominant  cold phase. If, on  the other hand,
these filaments  are destroyed, the  histograms will be  dominated by
the hot phase out to the halo Virial radius.

\section{Properties of diffuse gas accretion}

We  computed the  accretion-weighted  PDFs for  several hundred  haloes
spanning  dark matter masses between $ 10^{10}-10^{13} \Msun$  between $2\le z \le  5$. 
%We select haloes from the FOF catalogue at each redshift so as to keep all the most massive objects (for instance $\MDM \ge 10^{12} \Msun$ at z=2.5), which are also the rarest. At smaller masses, the number of haloes available is much larger (following the halo mass function of the simulation, i.e. a power law), and the selection is centered on a few discrete masses separated typically by 0.4 dex, down to $5 * 10^{9} \Msun$.
%Having such a discrete rather than continous mass spectrum for the lower mass haloes has no effect on the results since the objects are generally binned in 
We then  co-added (stacked)
these distributions for haloes of the  same mass range in order to produce
an ``average'' PDF  for a given mass scale. 
%Because of this stacking, we expect that having a discrete rather than continous mass spectrum for the lower mass haloes has no impact on the results. 
We use  these stacked PDFs to study the typical  temperature  and metallicity  distribution  of the  accretion
flow, and the  transition mass between the hot  dominated and the cold
dominated accretion regimes. Throughout the whole paper, we use exclusively a basis 10 logarithm, $\log_{10}$, and will therefore drop the subscript $_{10}$ in text, equations and figures.

%\begin{figure*}
%\begin{tabular}{c@{}c@{}c@{}c@{}c}
%&
%{\includegraphics[width=5.7cm]{fig1_collabel1.ps}}&
%{\includegraphics[width=5.7cm]{fig1_collabel2.ps}}&
%{\includegraphics[width=5.7cm]{fig1_collabel3.ps}}&
%\\
%
%{\includegraphics[height=5.7cm]{fig1_1x0.ps}} &
%{\includegraphics[height=5.7cm]{fig1_1x1.ps}}&
%{\includegraphics[height=5.7cm]{fig1_1x2.ps}}&
%{\includegraphics[height=5.7cm]{fig1_1x3.ps}}&
%{\includegraphics[height=5.7cm]{fig1_1x4.ps}} 
%\\
%{\includegraphics[height=5.7cm]{fig1_1x0.ps}} &
%{\includegraphics[height=5.7cm]{fig1_2x1.ps}}&
%{\includegraphics[height=5.7cm]{fig1_2x2.ps}}&
%{\includegraphics[height=5.7cm]{fig1_2x3.ps}}&
%{\includegraphics[height=5.7cm]{fig1_2x4.ps}} 
%\\
%{\includegraphics[height=5.7cm]{fig1_1x0.ps}} &
%{\includegraphics[height=5.7cm]{fig1_3x1.ps}}&
%{\includegraphics[height=5.7cm]{fig1_3x2.ps}}&
%{\includegraphics[height=5.7cm]{fig1_3x3.ps}}&
%{\includegraphics[height=5.7cm]{fig1_3x4.ps}} 
% \\
%&
%{\includegraphics[width=5.7cm]{fig1_5x1.ps}}&
%{\includegraphics[width=5.7cm]{fig1_5x1.ps}}&
%{\includegraphics[width=5.7cm]{fig1_5x1.ps}}&

%\end{tabular}
%\caption{Accretion--weighted PDFs for 3 mass bins $[2\times 10^{10} \Msun,
%  2\times 10^{11} \Msun, 2\times 10^{12} \Msun]$ (from left to right). {\em
%    Top:} z=4, radially averaged. {\em middle:} z=2.5, radially
%  averaged. {\em bottom:} z=2.5, $0.2 \Rv$. The numbered labels on the contours give the logarithm of the PDF. When the hot phase is well developed,
%  there is a clear bimodality in temperature and metallicity.}
%\label{f:bigDF4}
% done with ???
%\end{figure*}

\begin{figure*}
\begin{tabular}{c@{}c@{}c@{}c@{}c}
&
{\includegraphics[width=5.3cm]{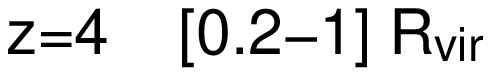}}&
{\includegraphics[width=5.3cm]{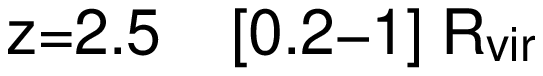}}&
{\includegraphics[width=5.3cm]{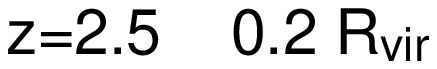}}&
\\
{\includegraphics[height=5.3cm]{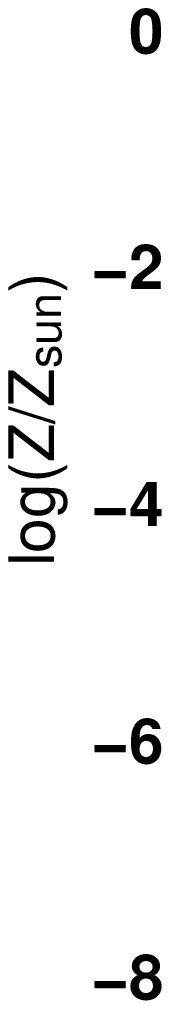}} &
%{\includegraphics[height=5.3cm,clip=true,bb=136 326 473 666]{./figs/new_TZmaps/z=4/8_stacked_DF4.ps}}&
%{\includegraphics[height=5.3cm,clip=true,bb=136 326 473 666]{./figs/new_TZmaps/z=2.5_avg/8_stacked_DF4.ps}}&
%{\includegraphics[height=5.3cm,clip=true,bb=136 326 473 666]{./figs/new_TZmaps/z=2.5_0.2Rv/8_stacked_DF4.ps}}&
{\includegraphics[height=5.3cm,clip=true,bb=136 326 473 666]{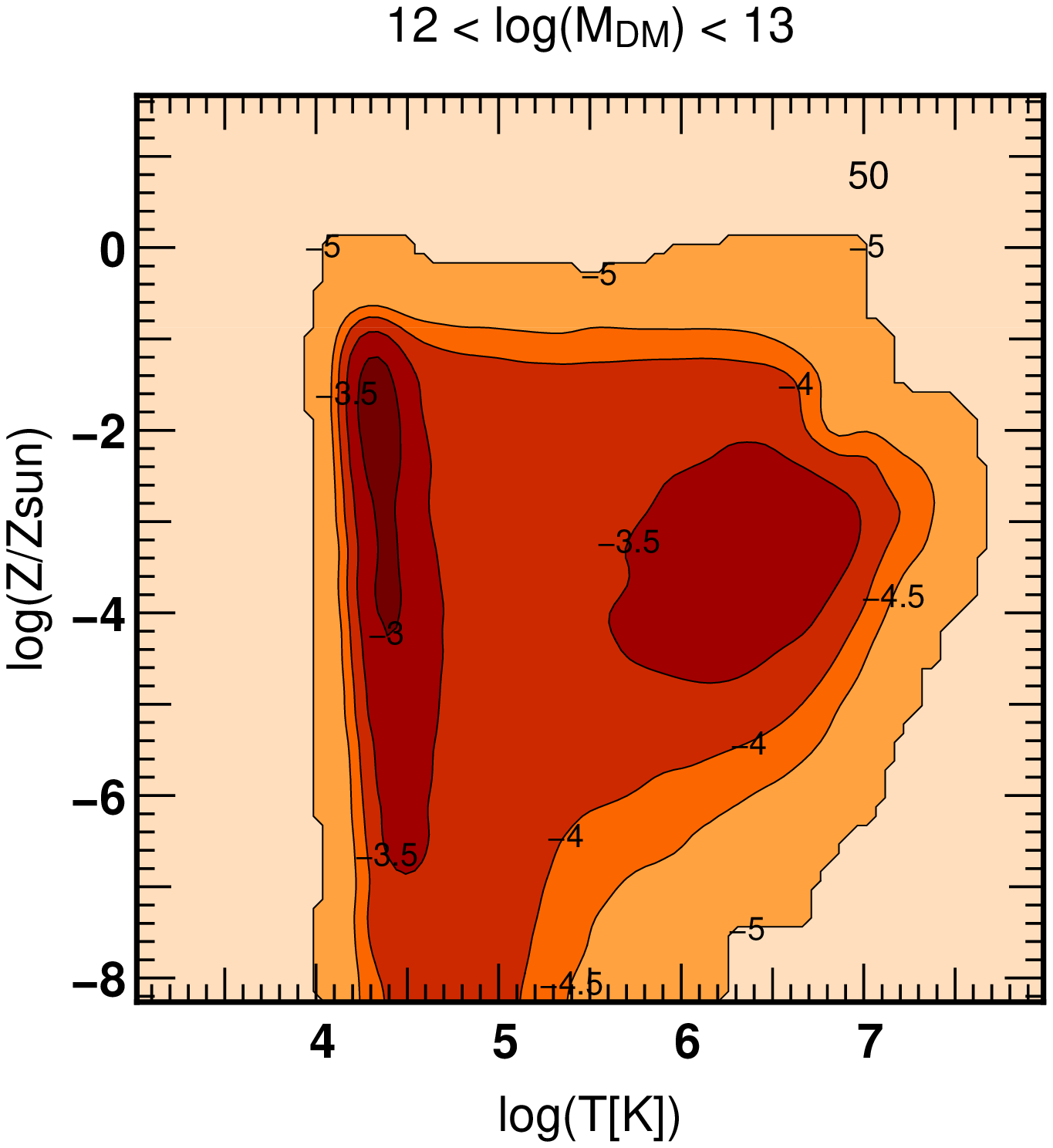}}&
{\includegraphics[height=5.3cm,clip=true,bb=136 326 473 666]{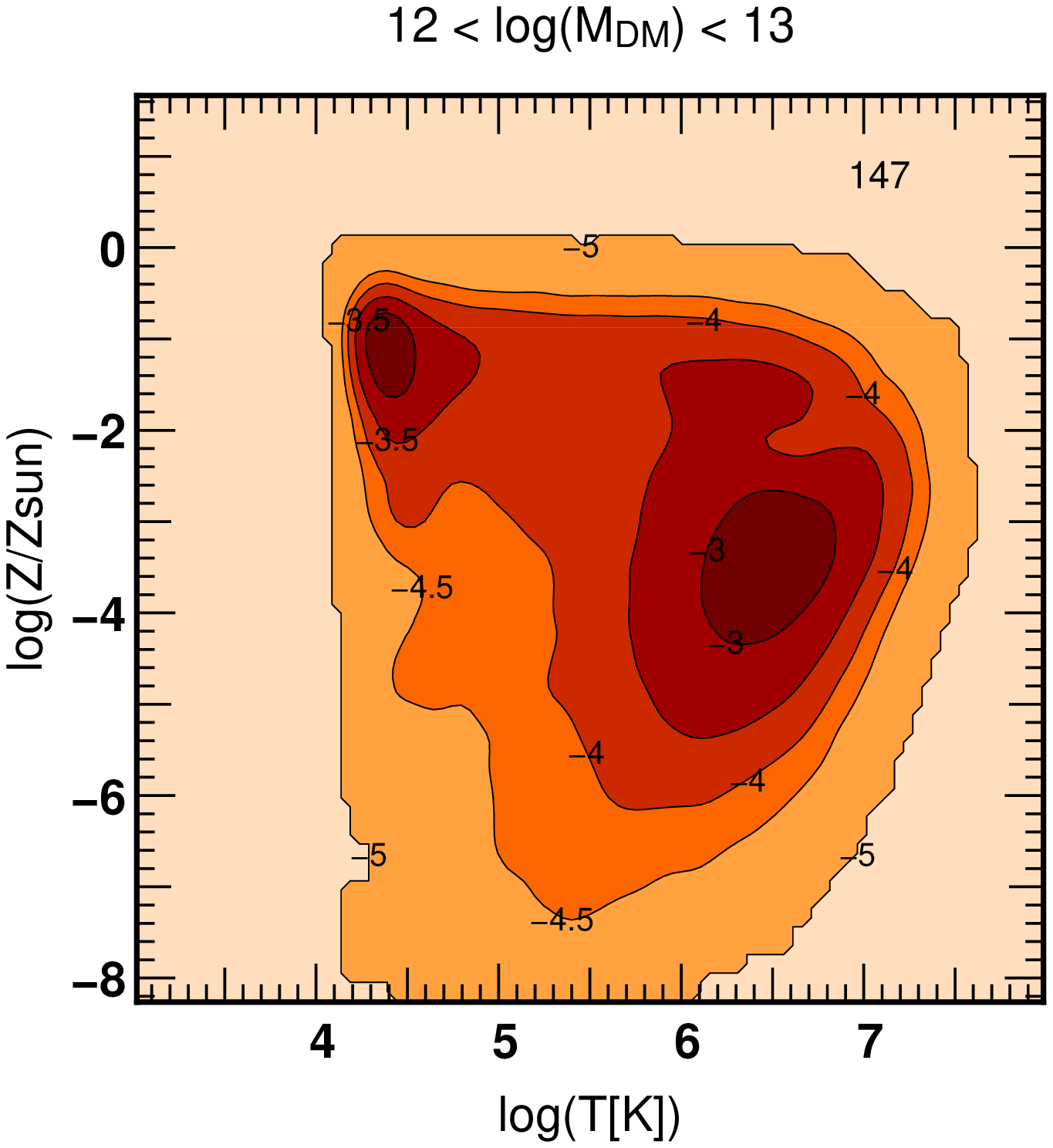}}&
{\includegraphics[height=5.3cm,clip=true,bb=136 326 473 666]{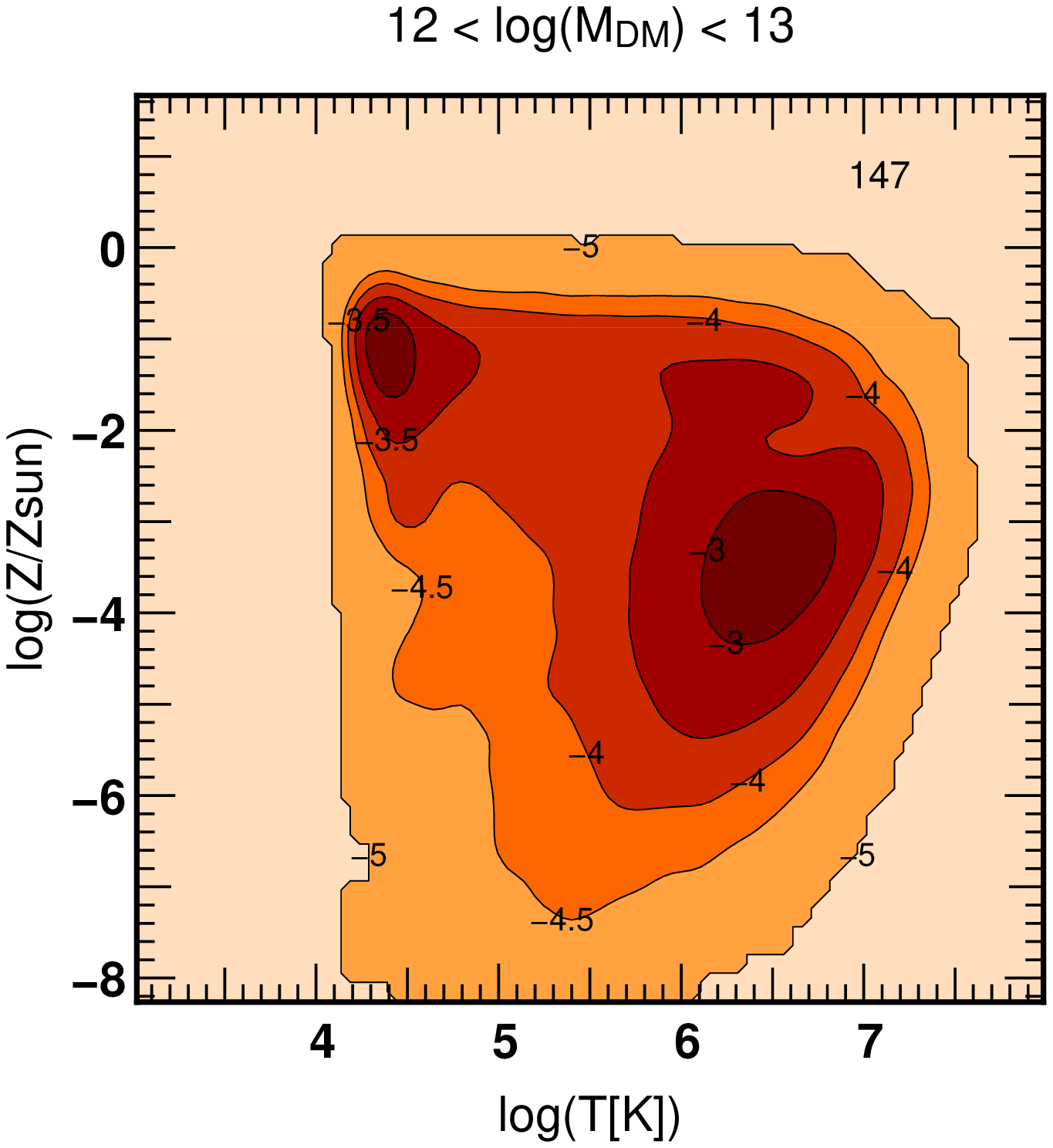}}&
{\includegraphics[height=5.3cm]{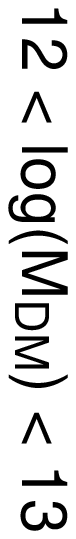}} 
\\
{\includegraphics[height=5.3cm]{fig1_1x0.ps}} &
{\includegraphics[height=5.3cm,clip=true,bb=136 326 473 666]{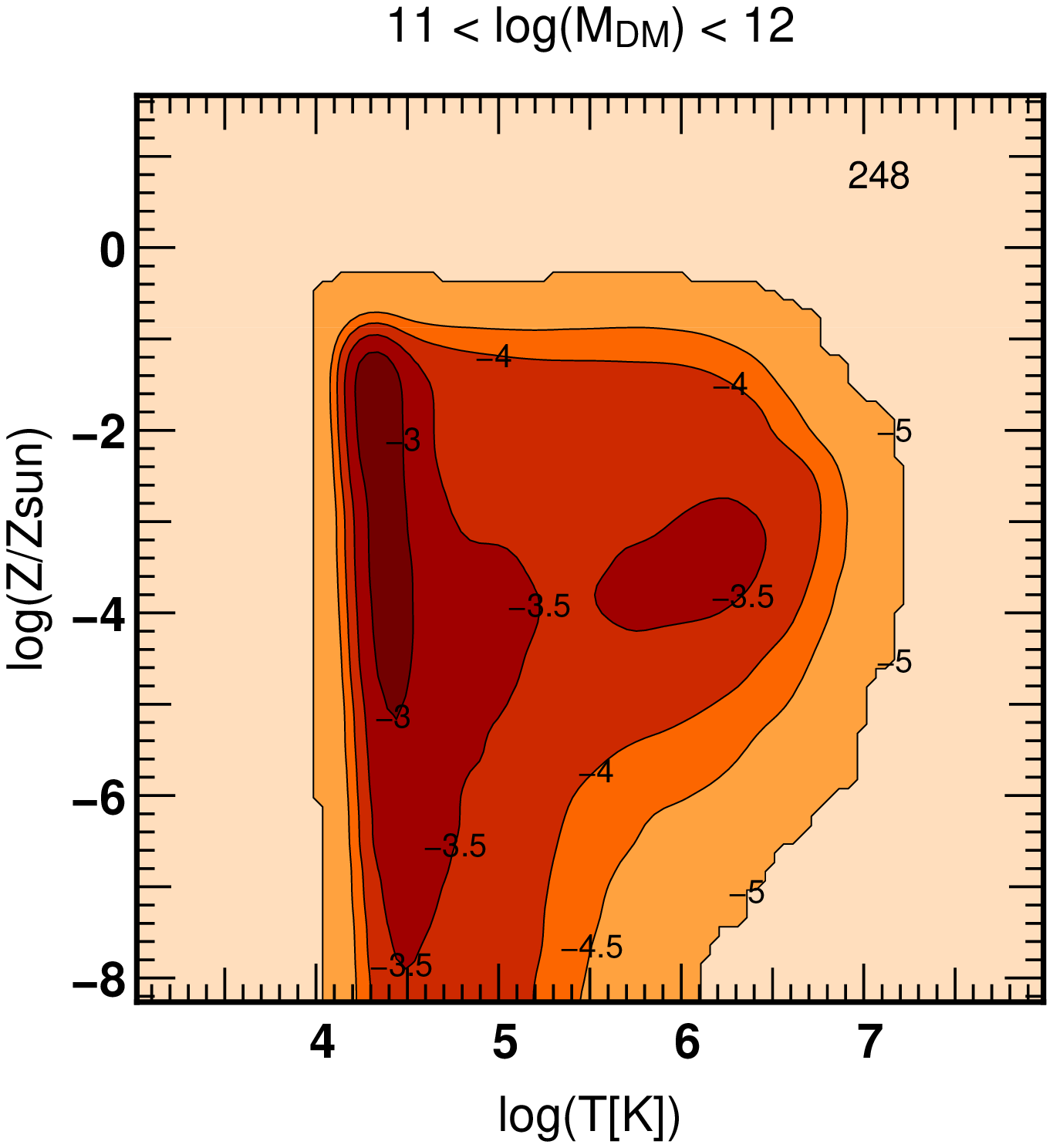}}&
{\includegraphics[height=5.3cm,clip=true,bb=136 326 473 666]{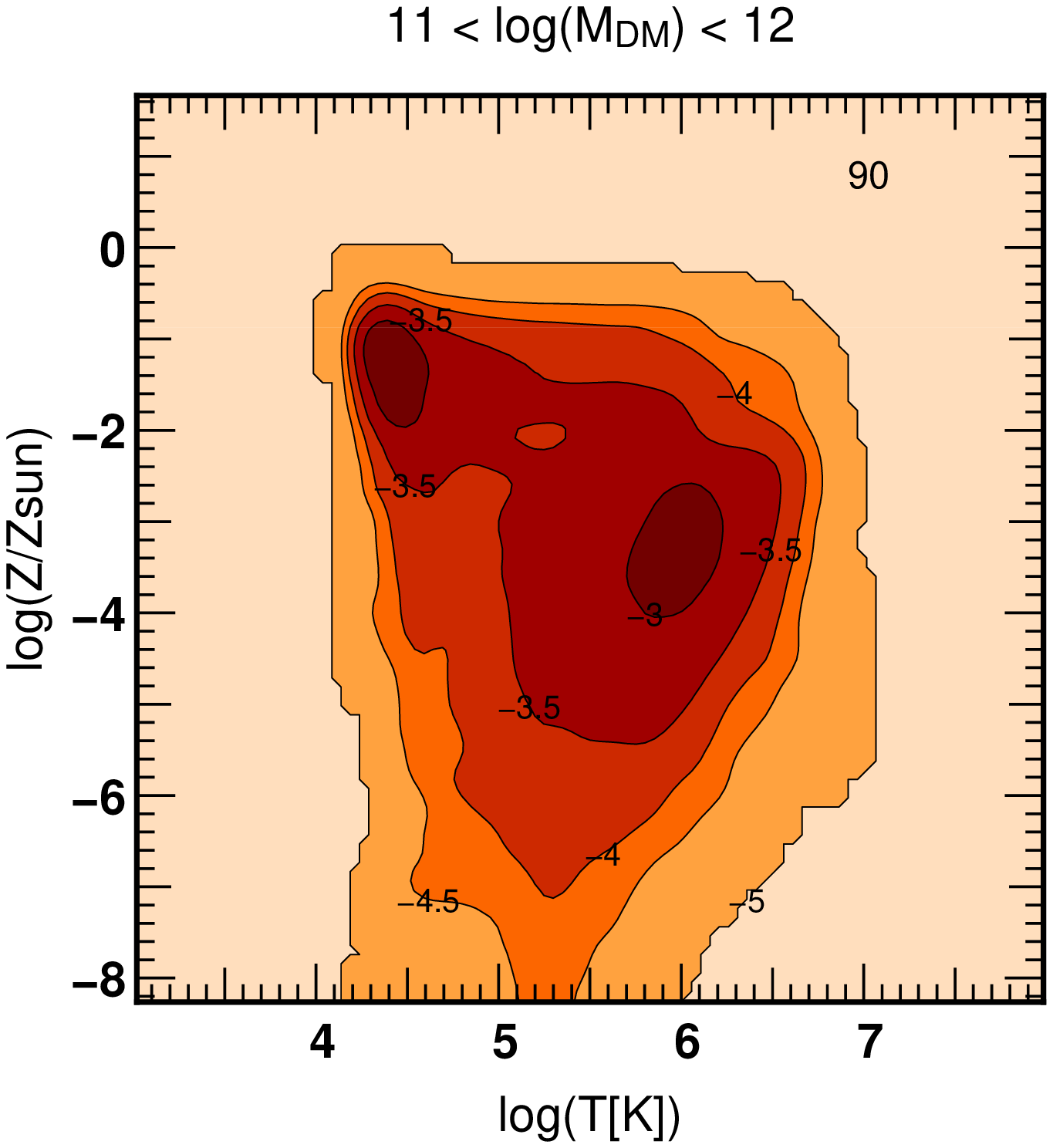}}&
{\includegraphics[height=5.3cm,clip=true,bb=136 326 473 666]{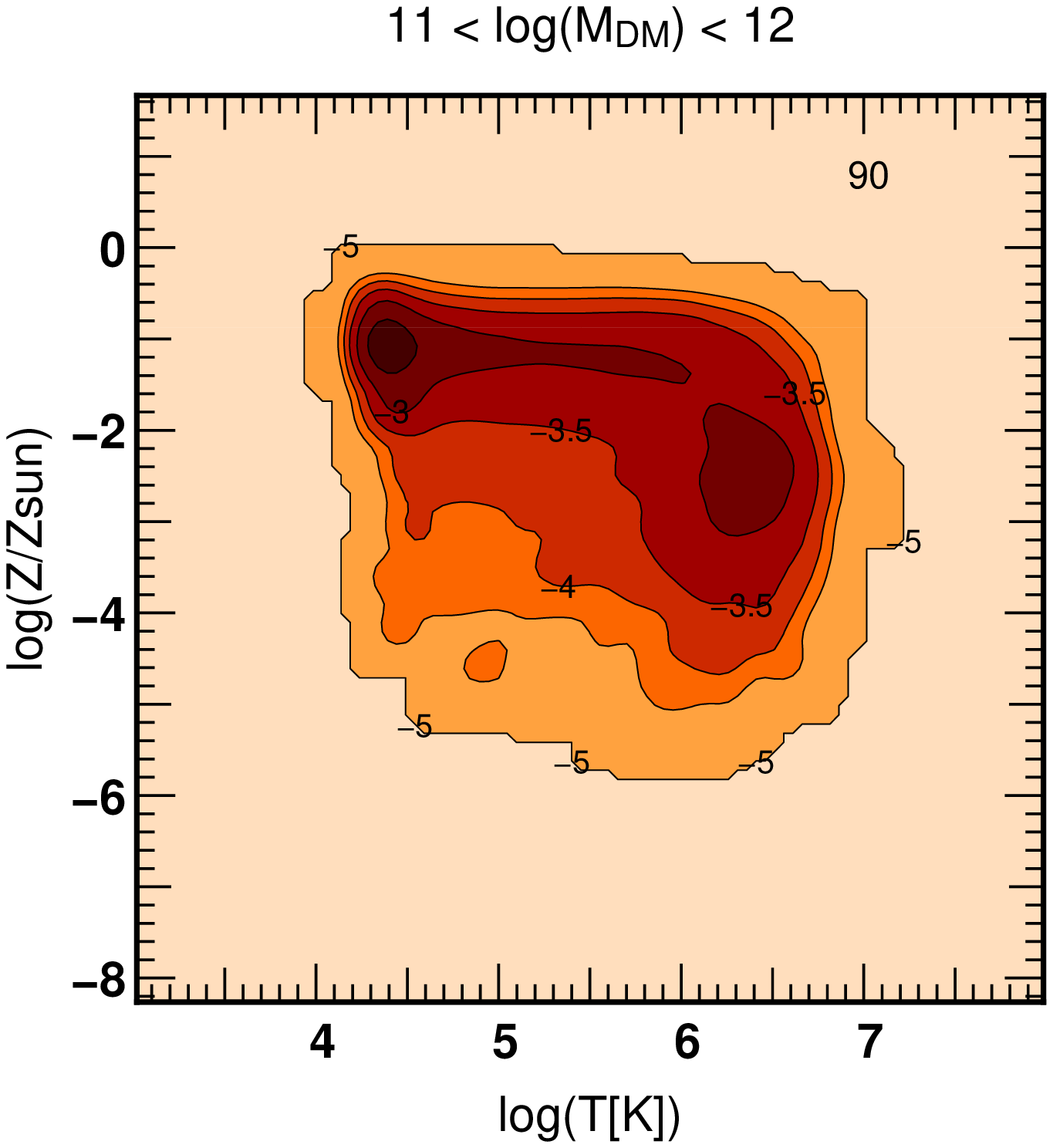}}&
{\includegraphics[height=5.3cm]{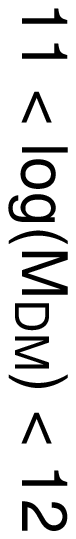}} 
\\
{\includegraphics[height=5.3cm]{fig1_1x0.ps}} &
{\includegraphics[height=5.3cm,clip=true,bb=136 326 473 666]{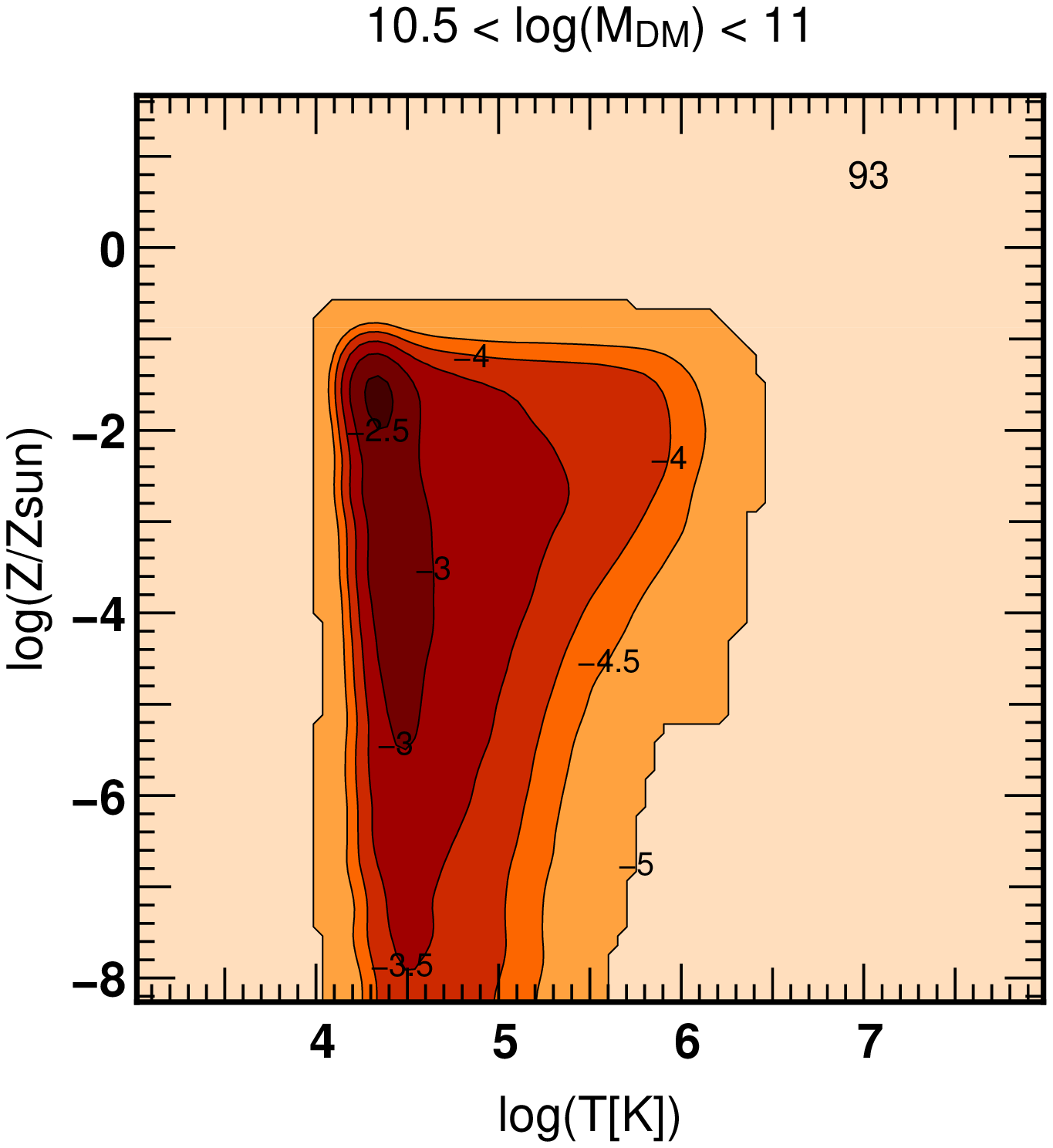}}&
{\includegraphics[height=5.3cm,clip=true,bb=136 326 473 666]{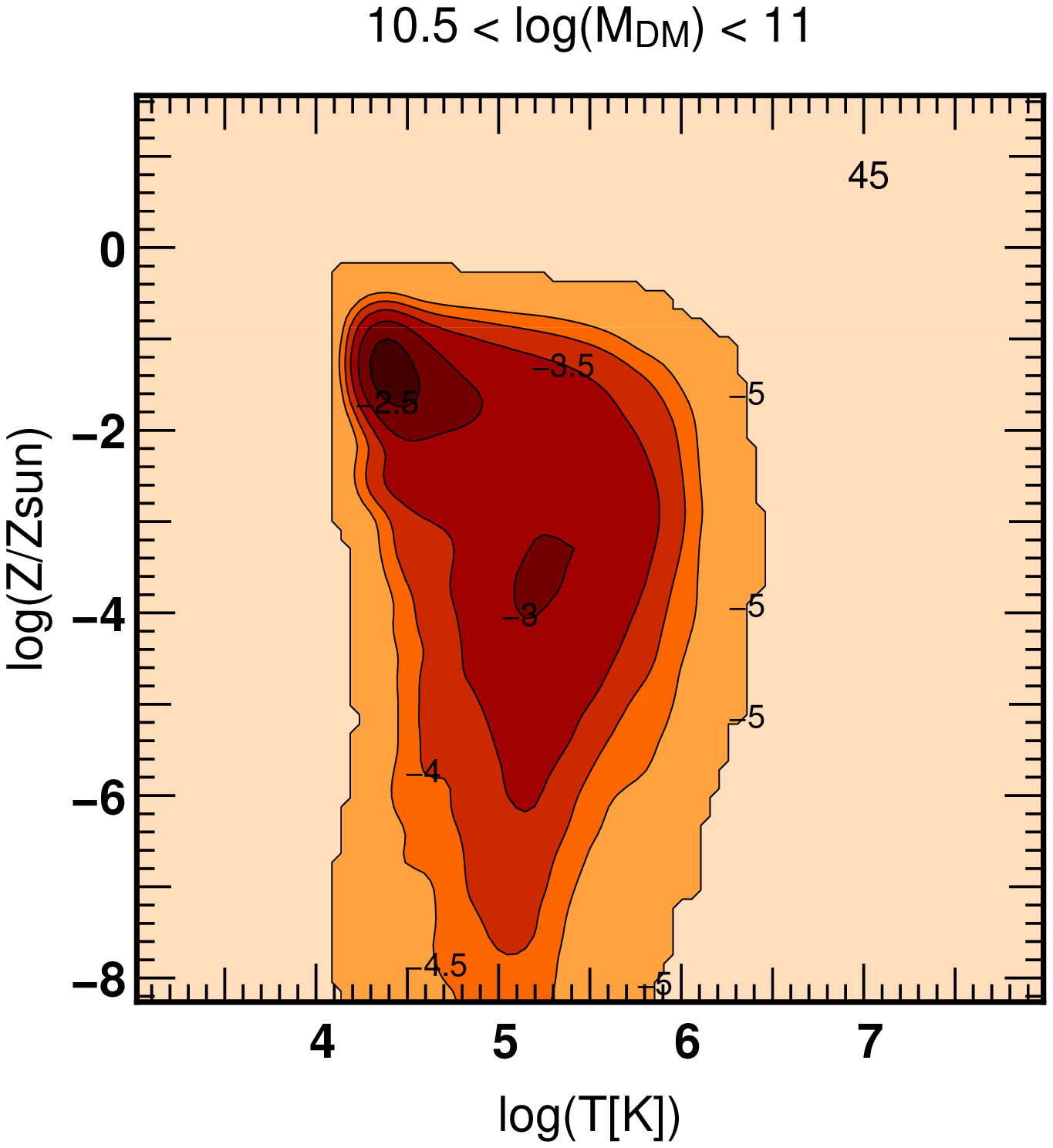}}&
{\includegraphics[height=5.3cm,clip=true,bb=136 326 473 666]{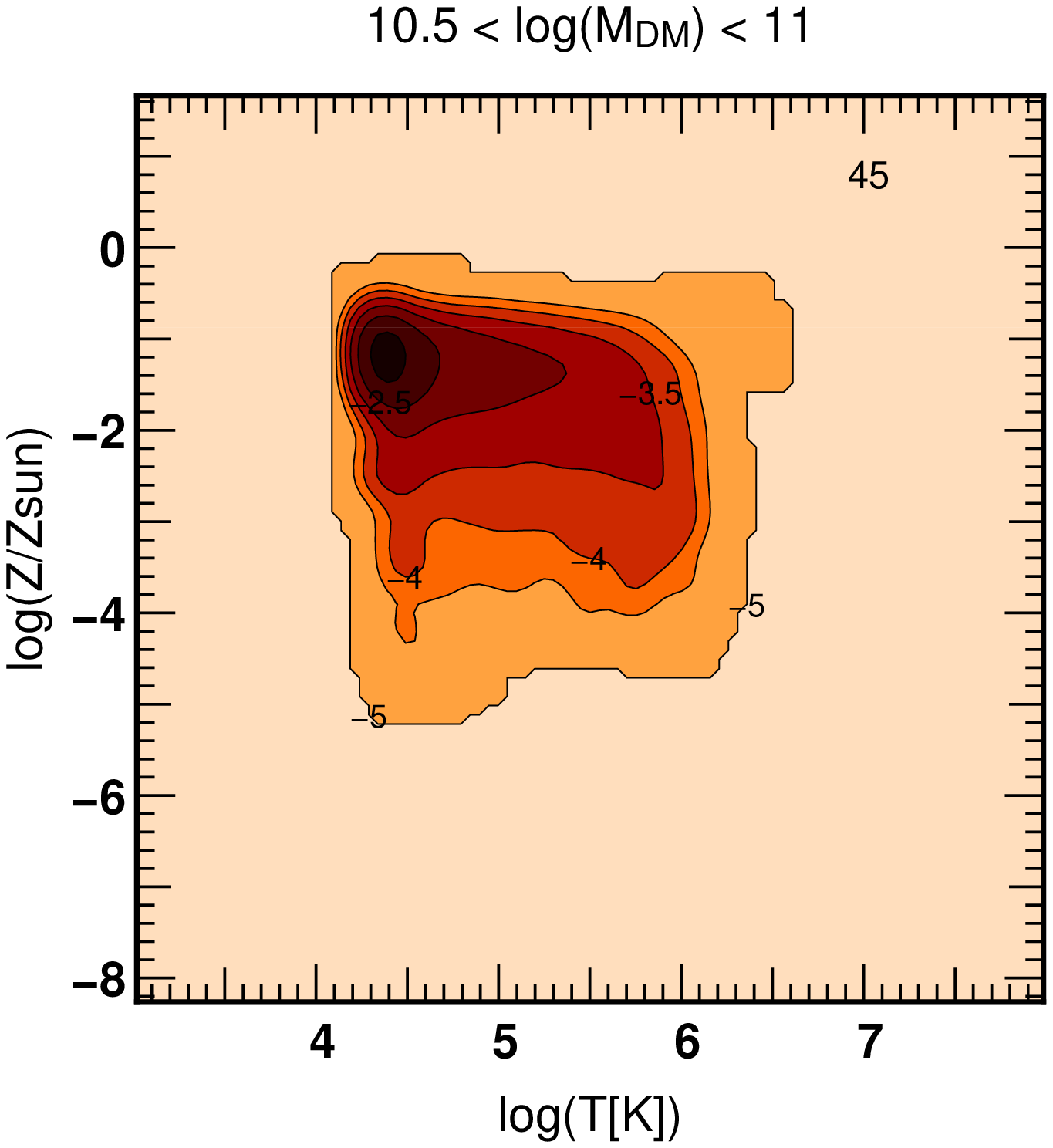}}&
{\includegraphics[height=5.3cm]{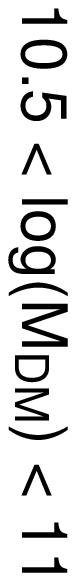}} 
\\
{\includegraphics[height=5.3cm]{fig1_1x0.ps}} &
{\includegraphics[height=5.3cm,clip=true,bb=136 326 473 666]{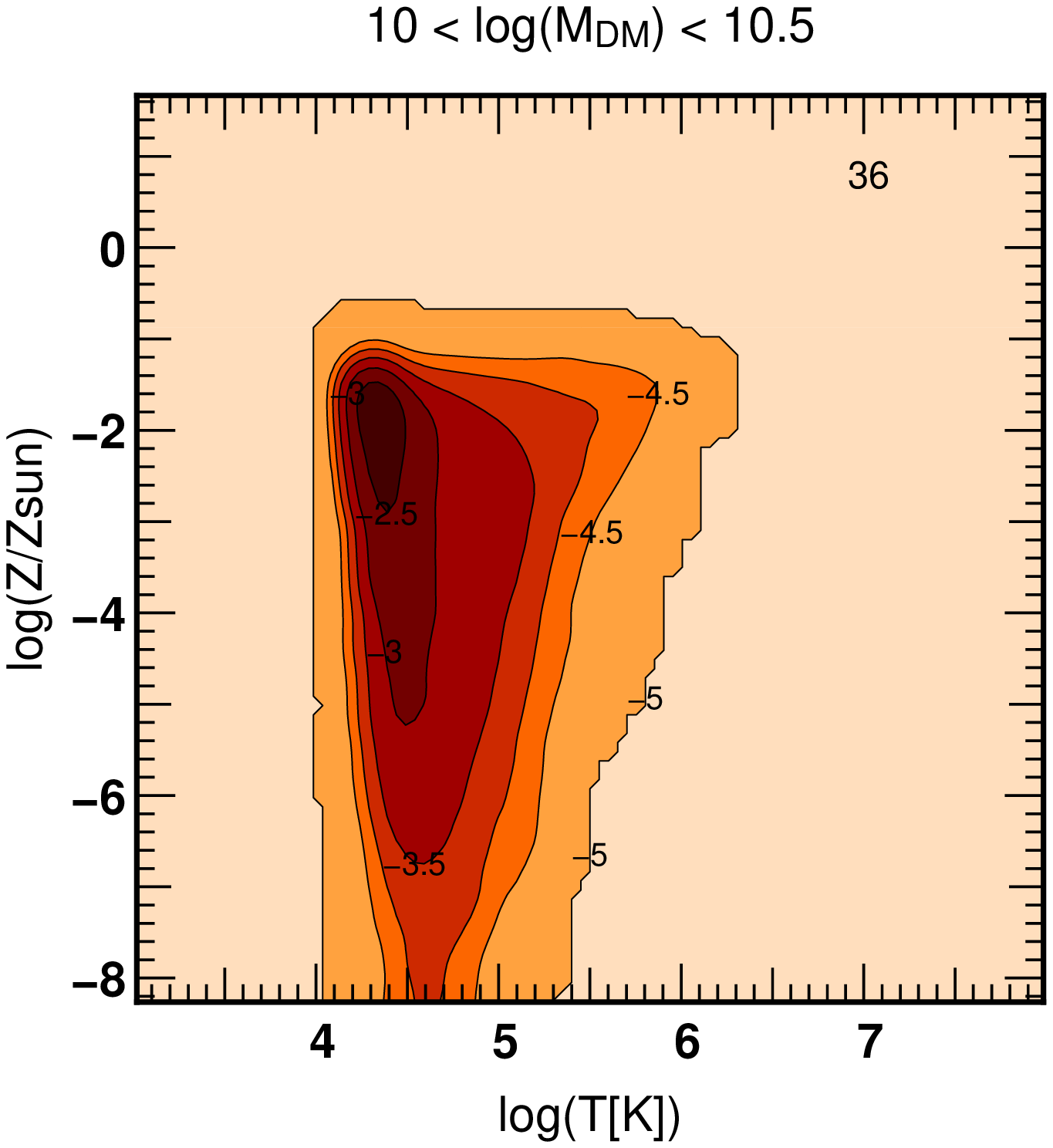}}&
{\includegraphics[height=5.3cm,clip=true,bb=136 326 473 666]{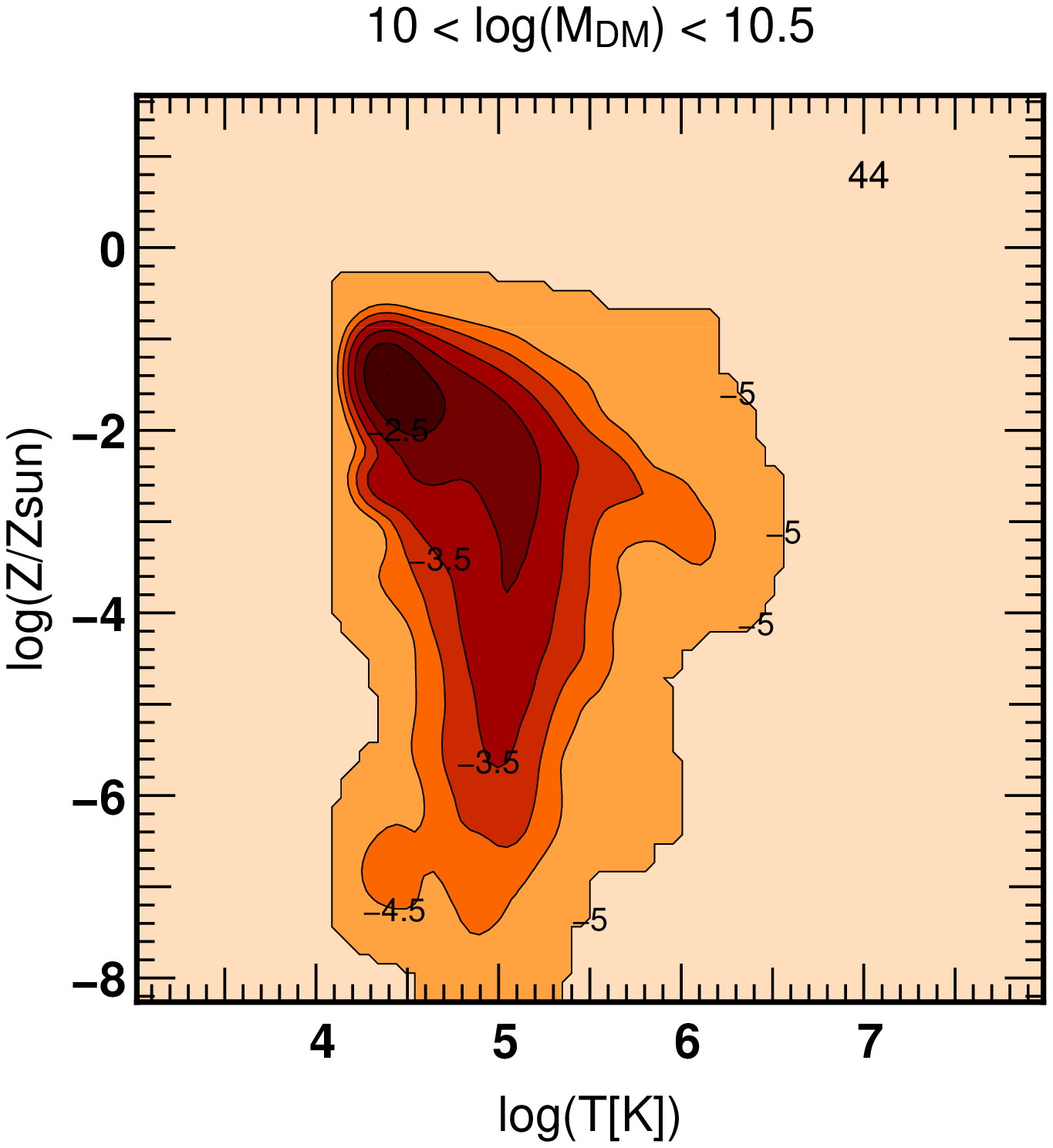}}&
{\includegraphics[height=5.3cm,clip=true,bb=136 326 473 666]{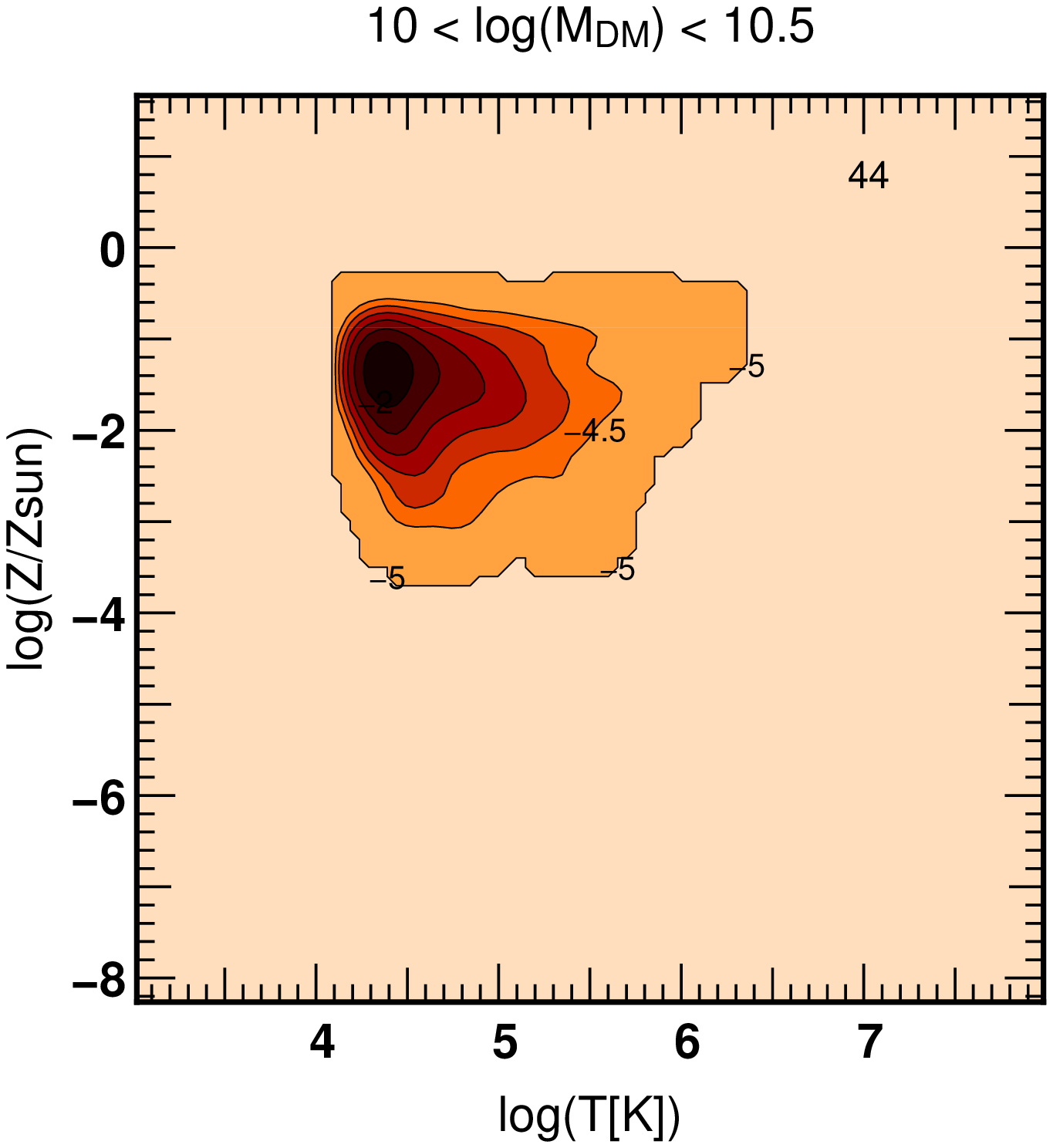}}&
{\includegraphics[height=5.3cm]{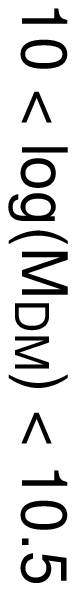}} 
\\
&
{\includegraphics[width=5.3cm]{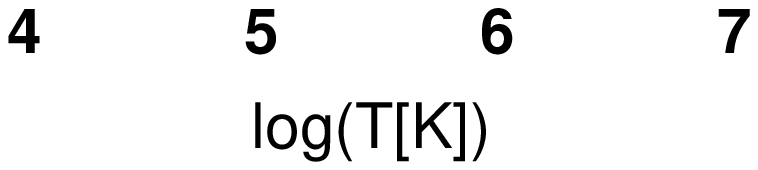}}&
{\includegraphics[width=5.3cm]{fig1_5x1.ps}}&
{\includegraphics[width=5.3cm]{fig1_5x1.ps}}&

\end{tabular}
\caption{Accretion--weighted PDFs for 4 mass bins from $10^{10}$ to $10^{13} \Msun$ (from bottom to top). {\em
    Left:} $z=4$, radially averaged. {\em Middle:} $z=2.5$, radially
  averaged. {\em Right:} $z=2.5$, $0.2$ $\Rv$. The numbered labels on the contours give the logarithm of the PDF. The number of galaxies in each mass bin is given in the top right cornerof each panel. When the hot phase is well developed,
  there is a clear bimodality in temperature and metallicity.}
\label{f:bigDF4}
% done with new TZmaps_for_revision1 or so...
\end{figure*}

\subsection{Bimodality in the temperature distribution}

\label{s:bimod}
\subsubsection{Hot and cold modes}
\label{s:bimod1}
The    left  and middle columns    of   Fig. \ref{f:bigDF4}    show    several
radially--averaged accretion--weighted  stacked--histograms (following Eq. (\ref{eq:defRMD})) for haloes
from $10^{10}$ to $10^{13} \Msun$ taken from the $z=4$ and $z=2.5$ snapshots of the simulation, respectively.  The right panel shows these histograms for the $0.2 \Rv$ spherical surface at $z=2.5$. The number of galaxies in each mass bin is given in the top right corner of each panel. These numbers reflect the halo sampling strategy and not the halo mass distribution of the simulation. For lisibility, the histograms have been  convolved with a smoothing kernel of about 0.2 dex FWHM in T and $Z$. We see that
the accretion pattern involves  two main distinct components:
\begin{enumerate}
\item{A cold component, the metallicity of which extends over several decades}
\item{A hot, relatively metal-poor component, the temperature and contribution of
which increases sharply with halo mass}
\end{enumerate}
As such, accretion itself is clearly bimodal in temperature. Indeed, at
any  halo mass, little  mass is  ever accreted  around $T  \approx 1-2.5
\times 10^5$ K. Instead, most of  the mass is accreted either below or
above this temperature.  As already noted by BD03 and \cite{keres05}, this
involves a link with the  physics of cooling: little mass will be accreted at
temperatures where  the cooling is efficient, since  gas cannot remain
at  this   temperature  for  very  long.   This   provides  a  natural
temperature threshold that allows us  to separate the cold and the hot
accretion modes, associated with  low and high mass haloes respectively.
The middle and right columns of Fig. \ref{f:bigDF4} show that
this bimodality in  temperature is also seen at $z=2.5$, in the whole halo as
well as at the inner halo. 

\begin{figure}
{\includegraphics[width=1.\linewidth,clip]{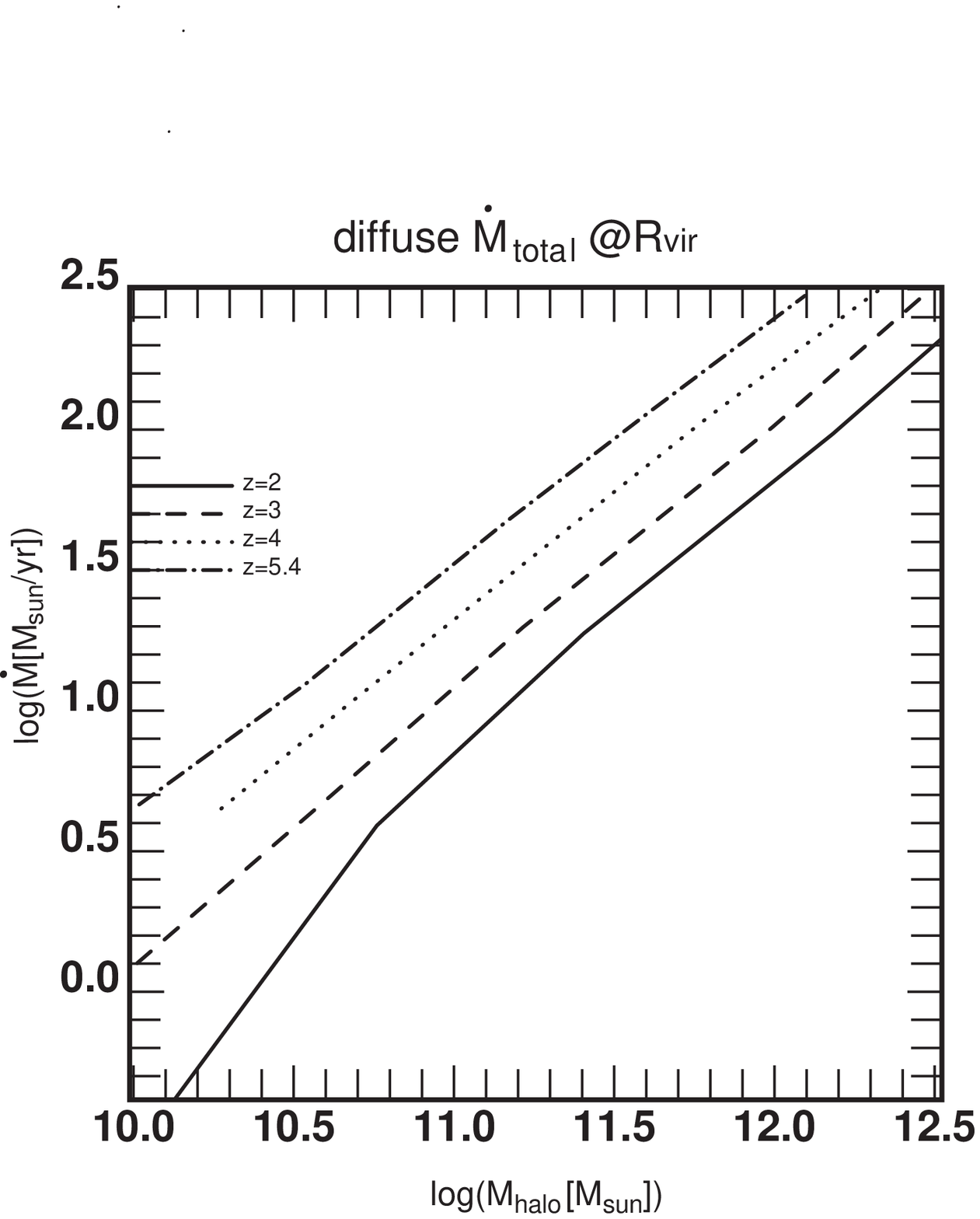}}
\caption{Diffuse gas accretion rate  measured at $\Rv$ versus dark matter halo mass at
various epochs between redshift 2 and 5.4}
% done with Mdottotal_paper.i
\label{f:Mdottotaldiffuse}
\end{figure}

A good criterion for
establishing  the  existence of  a  well-developed  hot  phase could be  to
require  the existence of  a saddle  point in  the accretion--weighted
PDFs of Fig. \ref{f:bigDF4},  as  a local
minimum in temperature and a local maximum in metallicity.  According to
this criterion,  the hot  phase of the lowest mass bin haloes is not  well-developed at any redshift between
$z=$  2--4. 
The  accretion-weighted  PDFs  can  be normalized  using  the  total
diffuse      gas     accretion      at     the      Virial     radius.
Fig. \ref{f:Mdottotaldiffuse}   shows   that   the  accretion   rate
decreases with cosmic  time for all masses and  increases with mass at
fixed redshift.   This evolution is  very similar to that  reported in
\cite{keres05,rasera06,guowhite07,neistein08},  in trend and  normalization. 

\subsubsection{Identification of the phases}
The cold component described in \Sec{bimod1} can itself be decomposed into two distinct sub-components with different behaviours, based on their metallicity. 
On one hand, the high-$Z$ part of the cold component is made of the surroundings of satellite galaxies, i.e. a region close enough to the satellite core to be chemically enriched, but far enough to escape the density truncation described in \Sec{trunc}. This can be clearly seen on \Fig{space}. This component is seen at all redshifts.
Conversely, the low-metallicity tail ($Z \le 10^{-3}$ $\Zsun$) of the cold component is very prominent at $z=4$ but disappears at $z=2.5$ in the inner and outer halo. It can be identified with the dense, cold, metal poor gas filaments seen in Fig. \ref{f:space}. This is best seen at $z=4$, where the filaments are better defined. Their density is somehow intermediate between the galaxy discs and the background. The left panel of Fig. \ref{f:space} clearly shows a cold {\em metal-poor} filament tunneling down all the way from almost 2 $\Rv$ to the central galaxy's disc, while the cold {\em metal-rich} denser phase makes up the surroundings of satellites. The hot phase, with intermediate density and metallicity, is distributed in a large bubble, of radius smaller than $\Rv$ at $z=4$ but significantly larger than $\Rv$ at later times. 

%Having a discrete rather than continous mass spectrum for the lower mass objects has no effect on the results presented in 
%The  accretion-weighted  PDFs  can  be normalized  using  the  total
%diffuse      gas     accretion      at     the      Virial     radius.
%Figure~\ref{f:Mdottotaldiffuse}   shows   that   the  accretion   rate
%decreases with cosmic  time for all masses and  increases with mass at
%fixed redshift.   This evolution is  very similar to that  reported in
%\cite{keres05,rasera06,guowhite07,neistein08},  in trend and  normalization.  
%For
%comparison  we plot  an  arbitrary  slope 1  curve,  which is  clearly
%steeper than  our measurements.  This means that  the specific diffuse
%gas accretion  rate ($\Mdot/M$)  {\em decreases} with  increasing halo
%mass. \pier{I DON'T UNDERSTAND THIS LAST POINT GIVEN THE PLOT.}

\subsection{Metallicity of the hot mode}
\label{s:zfit}

\subsubsection{The heterogeneous hot phase}

The top  panels of Fig. \ref{f:bigDF4} show that  the metallicity of
the  hot and  cold  phase can  differ by  up  to 3  decades. 
This  gap
involves a  huge difference  in the  ability of the  gas to  cool down
radiatively, further aggravated by the relative densities of the two
phases.  This bimodal metallicity  distribution is  also seen  at the
inner   halo.  This   has   important   consequences   for
models of galaxy formation and evolution,  and highlights the necessity  of treating the gas as being composed of two phases with widely different metallicities in SAMs.  Moreover, at a given
temperature, the large spread of the PDF in metallicity shows that the gas is
not well mixed. Indeed, a perfectly well mixed accreted gas would show up as a
narrow peak in metallicity. This non-homogeneity is not a result of the stacking of the PDFs of haloes with different metallicities, but is already seen in individual haloes, although in a noisier fashion. Neither is it the result of stacking the PDFs of several shells of different radii: the top right panel shows that the hot phase metallicity distribution is already 2-3 dex wide in a single shell at the inner halo for the most massive halos.

\subsubsection{Accretion-rate weighted metallicity}

Having acknowledged the chemical heterogeneity of the hot phase and its possible implications, we now turn to defining the mass accretion-rate weighted metallicity $\Zhot$ of the hot phase, as the metallicity the hot gas would have if it was perfectly well mixed.
\begin{equation}
\Zhot(r)=\frac{\int_{T_0}^{\infty} \mdot(r,T,Z) Z \, \intd T \intd Z}{\int_{T_0}^{\infty} \mdot(r,T,Z) \, \intd T \intd Z} \, .
\end{equation}
Fig. \ref{f:Zhot} shows the variation of $\Zhot$ with respect to radius for several mass bins at $z=2$, and for the largest mass bin at $z=3$.

\subsubsection{Metallicity profile}

%We also find that $\Zhot$ depends on the  distance to the halo  centre. Fig. \ref{f:Zhot} shows the evolution
%with radius of $\Zhot$, computed as:
%\begin{equation}
%\log(\Zhot(r))=\frac{\displaystyle \int_{T_0}^{\infty} \mdot(r,T,Z) \log(Z)\, \intd
%  T}{\displaystyle \int_{T_0}^{\infty} \mdot(r,T,Z) \, \intd T} \, , 
%\label{eq:Zhot_log}
%\end{equation}
%with $T_0=2.5 * 10^5$ K. 
For all mass masses, $\Zhot$ increases sharply towards the halo center, and all the haloes have a similar central metallicity $\Zhot(0.1\, \Rv)=0.1 \, \Zsun$. However, the slope of the metallicity profile appears to depend strongly on mass, and the outskirts of the more massive haloes are found to be more metal-rich than their low mass counterparts. It is not clear at this stage of the analysis of the simulation if this trend can be explained by metal-rich winds from the central galaxy alone or stripping and dilution of the metal-rich disks of infalling satellites. It is also likely that winds in infalling satellites {\em} and stripping work together to enrich the halo gas: winds deposit metals out of the plane of the satellite's disk, and hence facilitate the stripping of these metals from the satellite.

We also show that these metallicity profiles are well fitted between $r/\Rv=0.1-1$ and $\MDM=10^{10}-10^{13}\Msun$ by the simple logarithmic law:
\begin{equation}
\Zhot(r)=a + b \, \log(r/\Rv) \, ,
\label{eq:Zrfit}
\end{equation} 
where the dependence with respect to mass is restricted to the coefficients  a and b given by:
\begin{center}
\begin{minipage}{0.8\linewidth}
\begin{eqnarray}
{ a} &=& -8.725+0.5 \log(\MDM) \, , \\
{ b} &=& -7.725 +0.5 \log(\MDM) \, .
\end{eqnarray}
\end{minipage}
\end{center}
The thick gray lines in Fig. \ref{f:Zhot} show this fit for $\MDM=10^{12.5} \, \Msun$ (upper line) and $\MDM=10^{11.5} \, \Msun$ (lower line).

%Note that it is expected that this log-averaging gives a smaller metallicity than a purely linear averaging, but it follows more closely the maximum of the PDF. 
%Hence, an approximation of $\Zcool$ independent of mass can be a good description of our results in the $10^{10}-10^{13} \Msun$ range. In this respect, we find that  the dependency of $\Zhot$ with  radius at z=2 is well fit by the following law:
%\begin{equation}
%Z/\Zs=-4-2.7 \log \left( \frac{R}{\Rv} \right) \, ,
%\label{eq:Zrfit}
%\end{equation}
%where $\Rv$ is the Virial radius as defined in Sec.~\ref{s:method}. 

%This fit  is mainly independent of  mass and redshift  in the $2\times
%10^{10}$ -  $2\times 10^{12} \Msun$  range and between redshift $2\le z \le 5$. The dependence with respect to mass is only implicit through $\Rv$.
%At fixed redshift z=2 the metallicity depends much more strongly on radius than on mass in the considered mass range. Hence, this fit captures most of the metallicity-radius dependence and can be considered as a good approximation of the measurements (in a $\pm 0.25 $ dex range) for masses between $10^{10}-10^{13} \Msun$.
%This fit can be useful for analytical investigations (see Sec.~\ref{s:DB06}) or for SAMs of galaxy formation.

\subsubsection{Evolution with redshift}

\begin{figure}
{\includegraphics[width=1.\linewidth,clip]{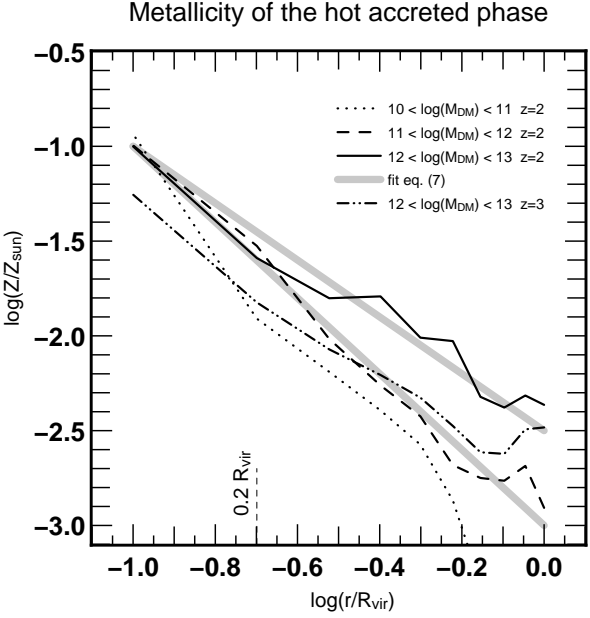}}
\caption{Metallicity  of the  hot accreted  gas as  a function  of the
normalized radius  $r/\Rv$ at $z=2$  for halo mass between  $10^{10}-
10^{13} \, \Msun$, and $z=3$ for the highest mass bin only. The thick gray lines  shows the analytical fit  of Eq. \ref{eq:Zrfit} for $\MDM=10^{12.5} \, \Msun$ (upper line) and $\MDM=10^{11.5} \, \Msun$ (lower line). The dependence with  respect to radius is very strong. While the profile gets steeper with decreasing mass, the innermost metallicity is almost constant.}
\label{f:Zhot}
% done with make_mass_weighted_Z_for_revision1.i
\end{figure}

Fig. \ref{f:Zhot} also shows the metallicity profile of the highest mass bin for $z=3$. %It is the only mass bin for which the hot phase is developed enough at all radii to allow for $\Zhot(r)$ measurements. 
The latter profile seems to be a mere downwards translation of the $z=2$ high mass profile, indicating that the enrichment rate $s= - \partial \log(\Zhot/\Zsun)/ \partial z \approx 0.2-0.3$ is almost the same at the inner halo and at the outer halo.
This compares rather well with the average  chemical enrichment
rate  $s=0.17$ found by \citet{delucia04}.
It is also in agreement with the central chemical enrichment rate of \cite{cora08}, who use a semi-analytical approach.

\subsubsection{X-ray clusters}

To roughly compare our profiles with metallicity measurements of X-ray clusters, we first need to extrapolate linearly the evolution of $\Zhot$ at the inner halo down to $z=0$. This gives $\log(\Zhot (0.1 \Rv, z=0)/\Zsun) \approx -0.4$ (assuming $s=0.3$), which is compatible with observations of \cite{vikhlinin05}. Indeed, they find $\log(Z/ \Zsun)=-0.2$ at 0.1 ${\rm R_{500}}$. The agreement further holds when comparing to the low-redshift part of the clusters of \cite{Balestra07} and \cite{maughan07}. Note however that the galaxy clusters studied in the latter papers have progenitors more massive than the most massive halos of our simulation, and thus expectedly overall higher metallicity. 
Indeed, our most massive halos are more likely to end up in groups of galaxies at $z=0$. Such objects also host hot X-ray emitting gas with about half-solar metallicity \citep{buote00}, which is in agreement with our extrapolation.
Our finding that the metallicity we measured at 0.1$\Rv$ does not depend on halo mass is further supported by the fact that observed galaxy clusters and galaxy groups have roughly similar central metallicity (while the metallicity at the outskirts can be very different).
Finally, the enrichment rates found for X-ray clusters by \cite{Balestra07} and \cite{maughan07} ($s \approx 0.3$ from $z=0$ to $z=1.3$ are well matched by the enrichment rate measured in the simulation  between $z=2-3$.
%However the drop in $\Zhot$ with increasing $R$ is steeper than in any of the cluster observations we reviewed here. 
%This problem is well known in the field and could be linked to the implementation of winds and the limited resolution of star formation and stripping physics.

%\begin{enumerate}
%\item{Bias in metallicity measurements of X-ray clusters: since the emissivity of the gas increases with temperature and metallicity, X-ray metallicity measurements are doubly biased towards hot and metallic regions. As a result, the measured metallicity is an overestimate of the average metallicity of the hot phase.}
%\item{Wind physics: the treatment of supernova feedback and stellar winds in the simulation could be insufficiently efficient at ejecting the produced metals up to the outer halo.}
%\item{Stripping physics}
%\item{The metallicity of the hot outer halo evolves radically between z=0-2. This is not in agreement with our current results, which show a very slow enrichment at the outer halo.}
%\end{enumerate}

\subsection{Two critical masses for diffuse gas accretion}

\begin{figure}
{\includegraphics[width=0.99\linewidth,clip]{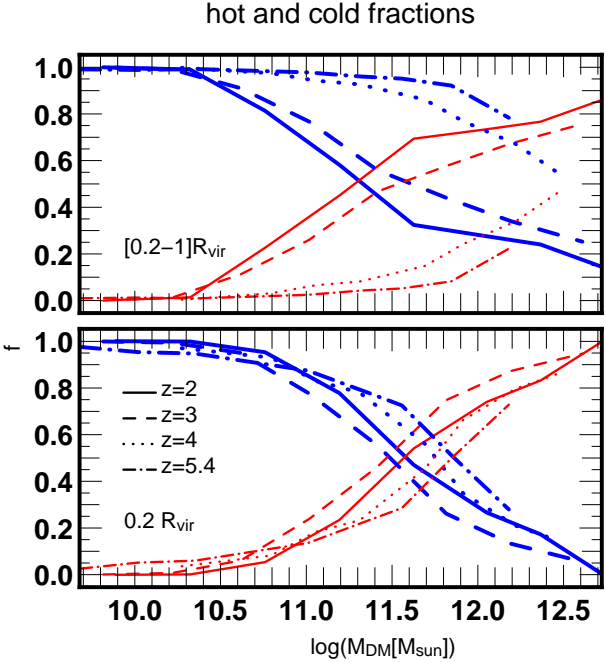}}
\caption{Evolution of the hot (thin line) and cold (thick lines) accreted gas mass fractions versus
$\MDM$   for   $z \in [2, 5.4]$.   Top:  $\langle \fcold \rangle$ and $\langle \fhot \rangle$  (integrated  from $0.1 \Rv$    down   to
0.2$\Rv$). Bottom: $\fcold$ and $\fhot$ on the 0.2$\Rv$ sphere. }
\label{f:hotcold}
\end{figure}

%To allow for comparison with  previous work \citep{keres05} we set the
%temperature  threshold  between  the hot  and cold  modes  at  $T_0=250\,000$
%K. 
Marginalizing  the accretion rate over  metallicity and integrating
over temperature on the hot and cold temperature domains
yields  the  hot and  cold  accretion  rate respectively.   Dividing by  the  total
accretion   rate  at   the  chosen   radius  gives  the
contributions of the hot and cold mode to the total accretion rate:
\begin{eqnarray}
\fcold(r) &=& \frac{1}{\Mdot(r)}\int_{T=0}^{T=T_0} \int_{Z=0}^{Z=\infty} \mdot(r,T,Z)  \intd T  \intd Z \, , \\
\fhot(r) &=& \frac{1}{\Mdot(r)} \int_{T=T_0}^{\infty} \int_{Z=0}^{Z=\infty} \mdot(r,T,Z) \intd T \intd Z\, \, ,
\end{eqnarray}
where $ \mdot(r,T,Z)$ is given by Eq. (\ref{eq:defMDRZ}). A similar 
expression involving $\langle \mdot(T,Z)\rangle$ and Eq. (\ref{eq:defRMD})
allows us to define $\langle\fcold\rangle  $ and $\langle \fhot \rangle$.
The  top panel  of Fig. \ref{f:hotcold} displays  the fractions
computed  from  the accretion-weighted  histograms  averaged over  the
entire  halo  (between  $0.2 \Rv$  and  $\Rv$).  The  bottom  panel   of  Fig.~\ref{f:hotcold}  shows  these  fractions
measured  at radius $0.2 \Rv$  (inner halo)  as a  function of
mass  for various  redshifts. 
A  common
feature  of  these plots  is  the  increasing  importance of  the  hot
accretion mode with increasing  mass, and the corresponding decreasing
contribution  of the  cold mode,  as could  be foreseen  from  the top
panels of  \Fig{bigDF4}. The mass at which  $\langle \fcold \rangle= \langle\fhot \rangle$ defines the
critical  mass  marking  the  transition  between  the  two  accretion
regimes. 

\begin{figure*}
\begin{tabular}{cc}
{\includegraphics[width=0.45\linewidth,clip]{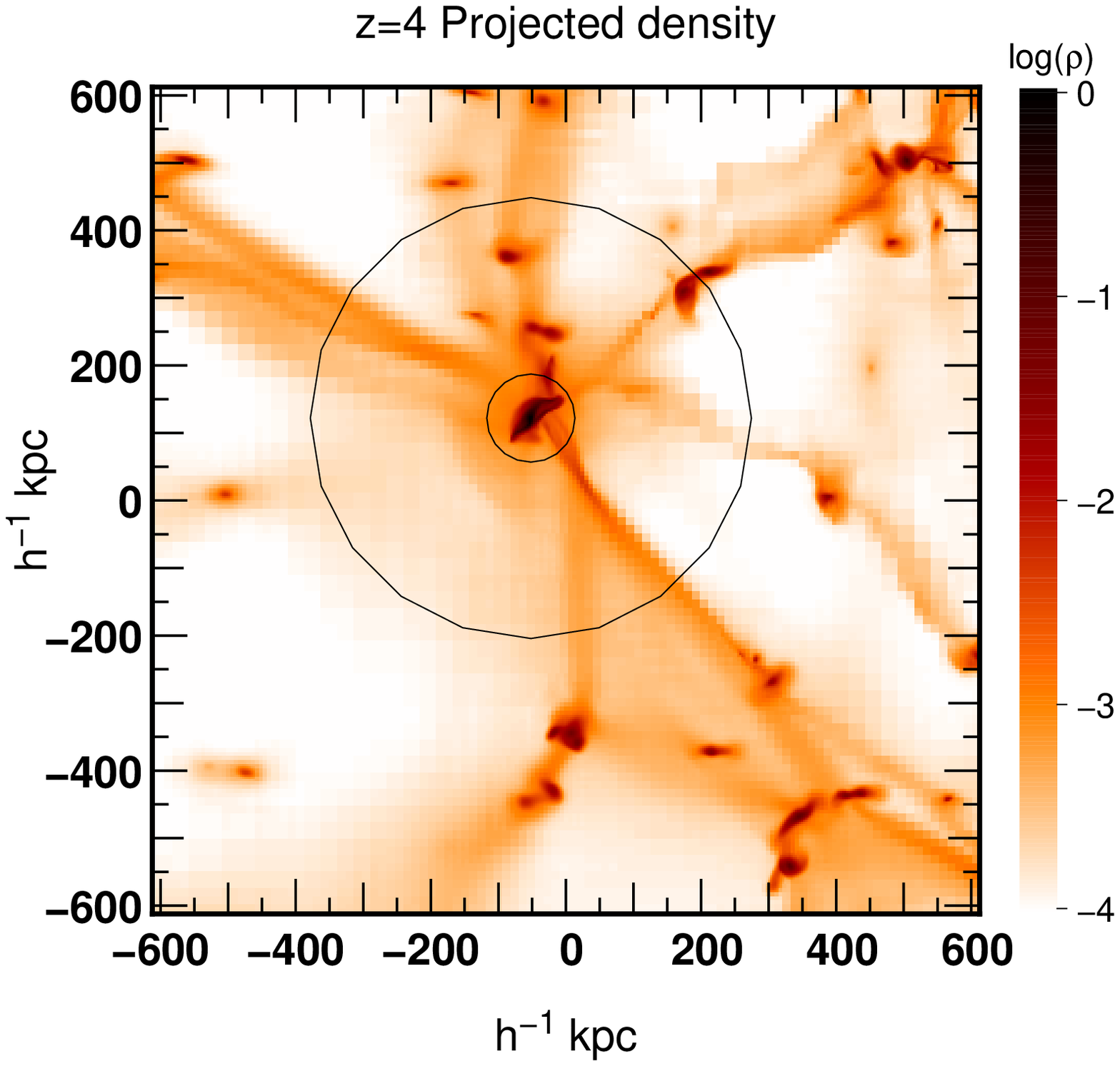}}
&
{\includegraphics[width=0.45\linewidth,clip]{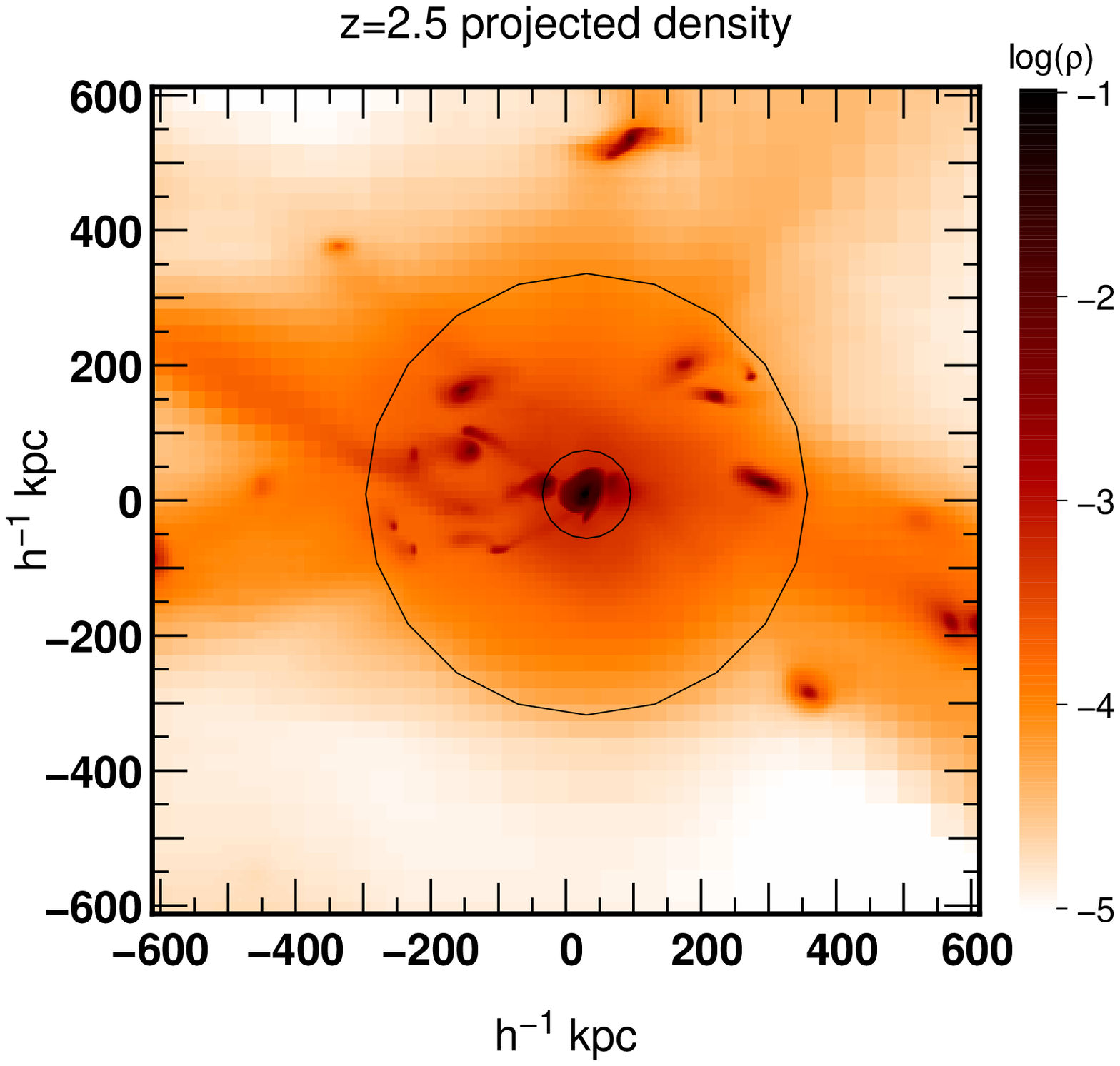}} \\
{\includegraphics[width=0.45\linewidth,clip]{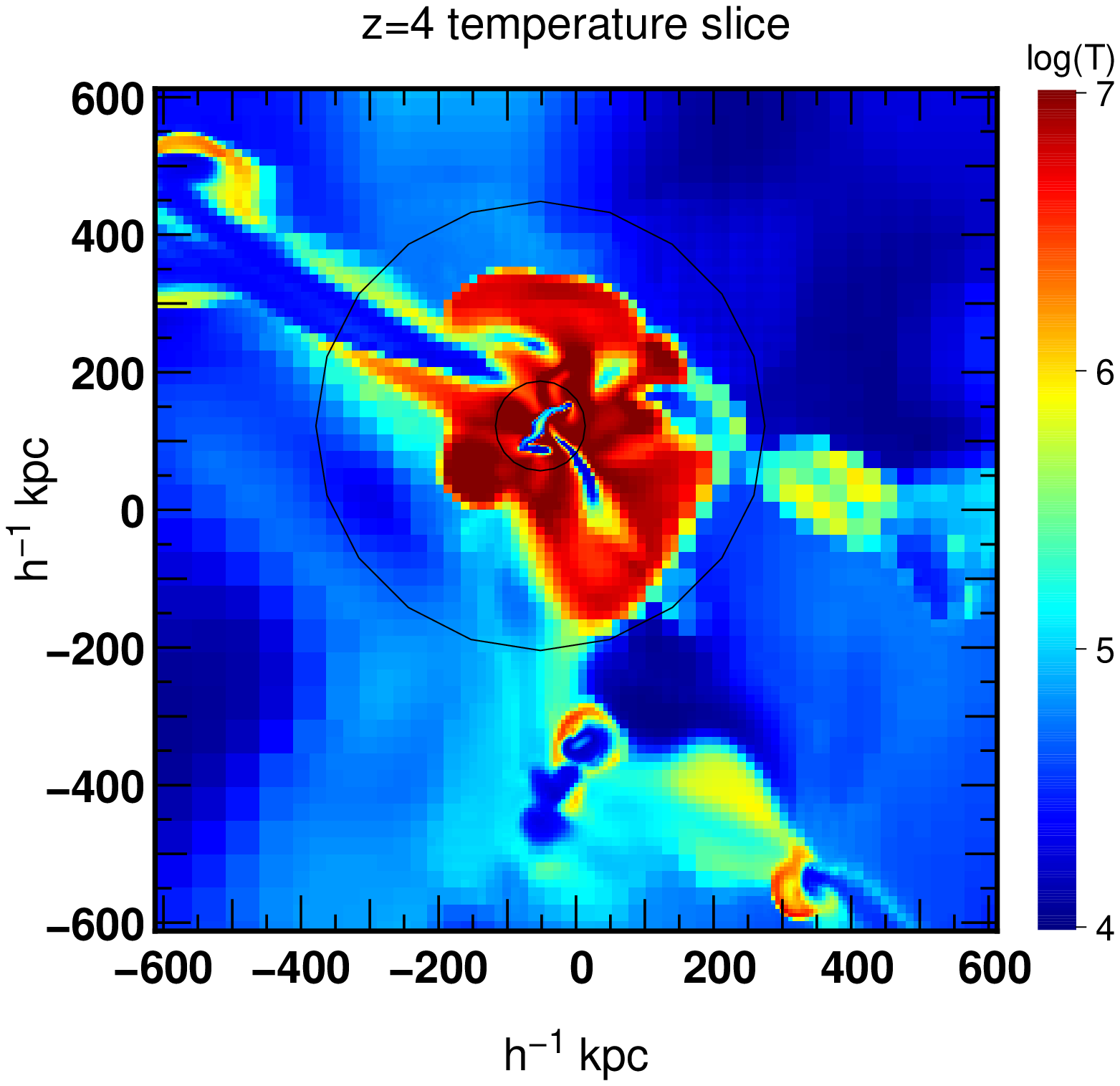}}
&
{\includegraphics[width=0.45\linewidth,clip]{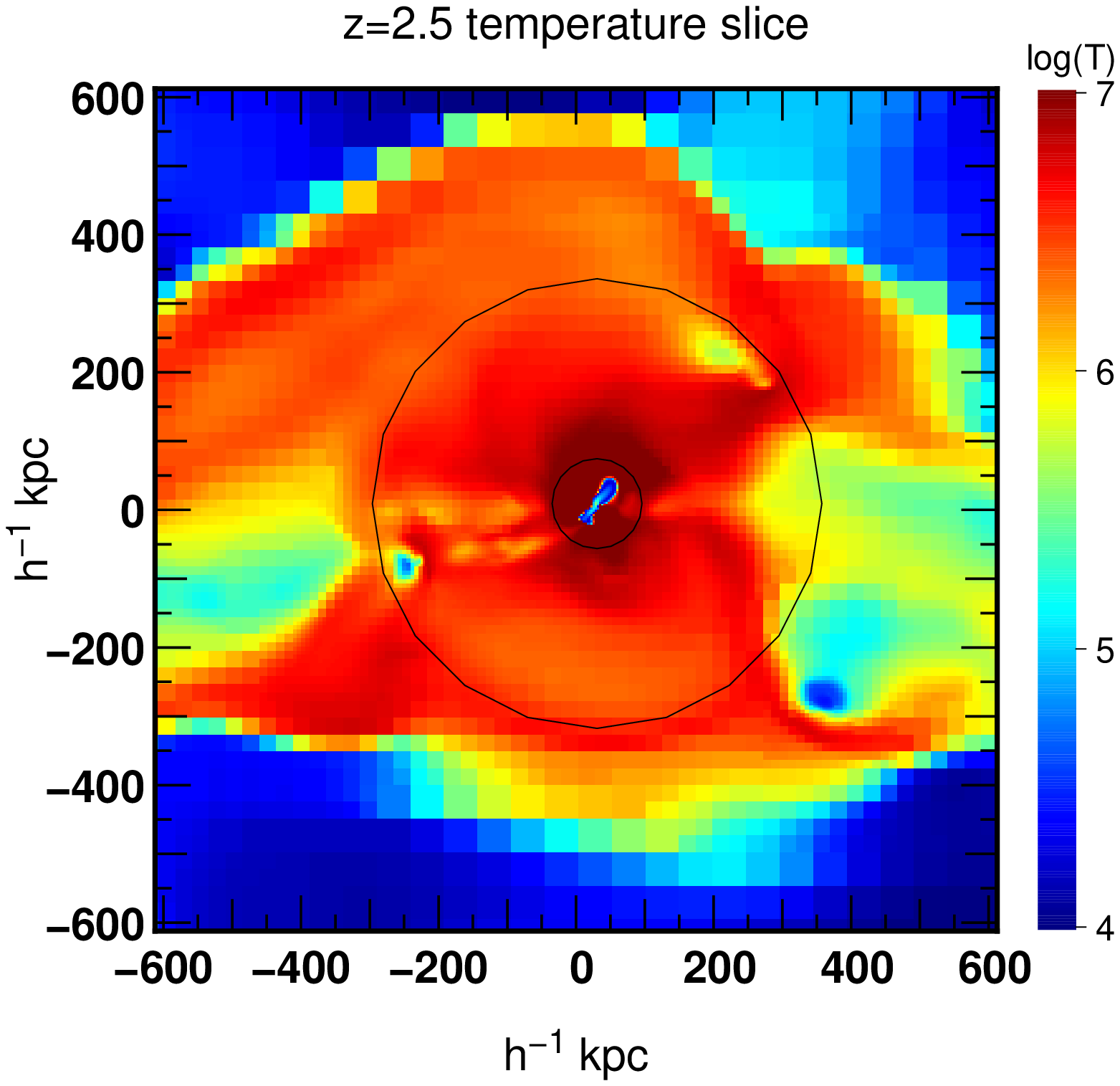}} \\

%{\includegraphics[width=0.45\linewidth,clip]{./figs/maps/040108/12_Zslice.eps}}&
{\includegraphics[width=0.45\linewidth,clip]{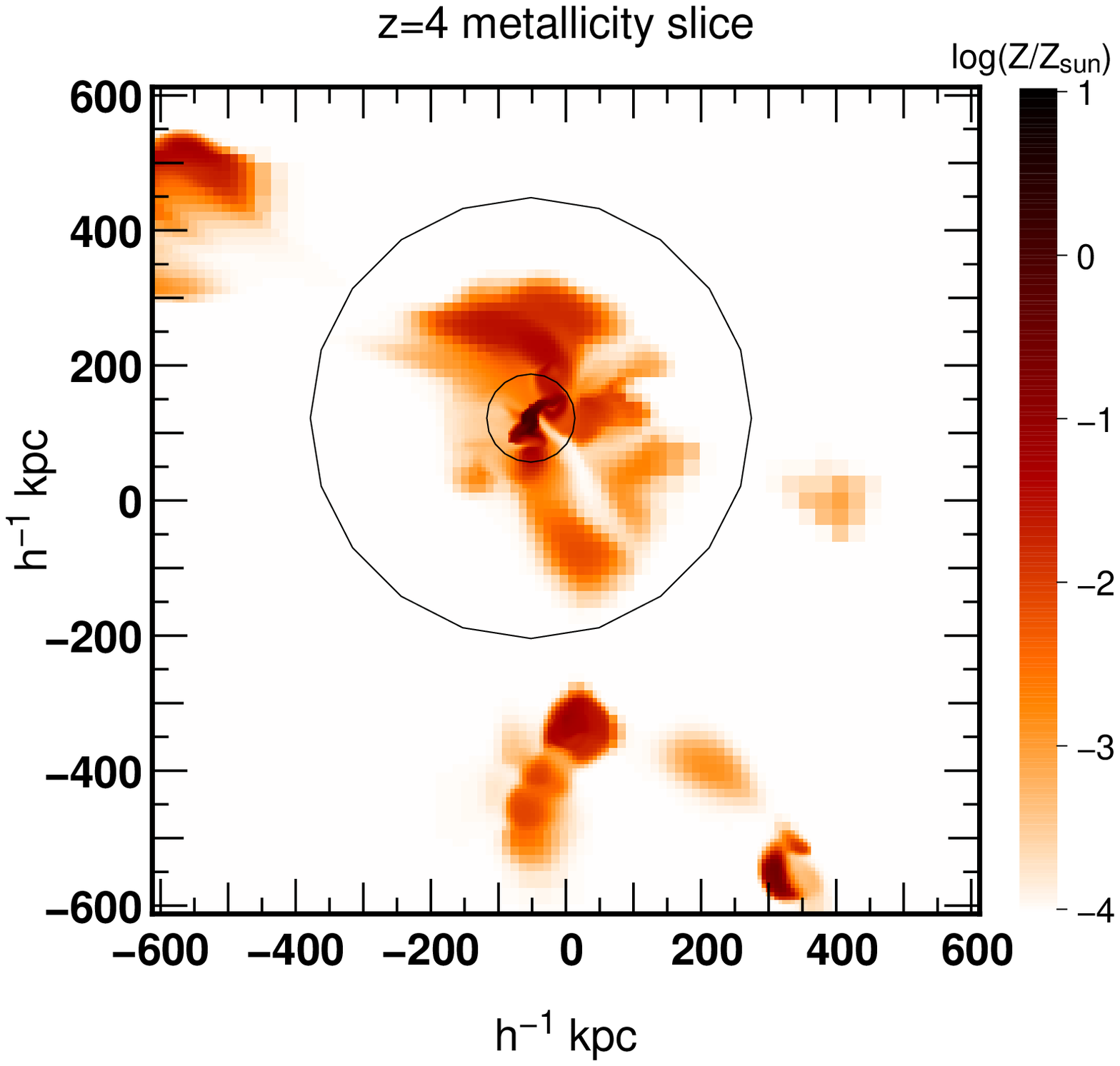}}&
%{\includegraphics[width=0.45\linewidth,clip]{./figs/maps/040108/81_Zslice.eps}}
{\includegraphics[width=0.45\linewidth,clip]{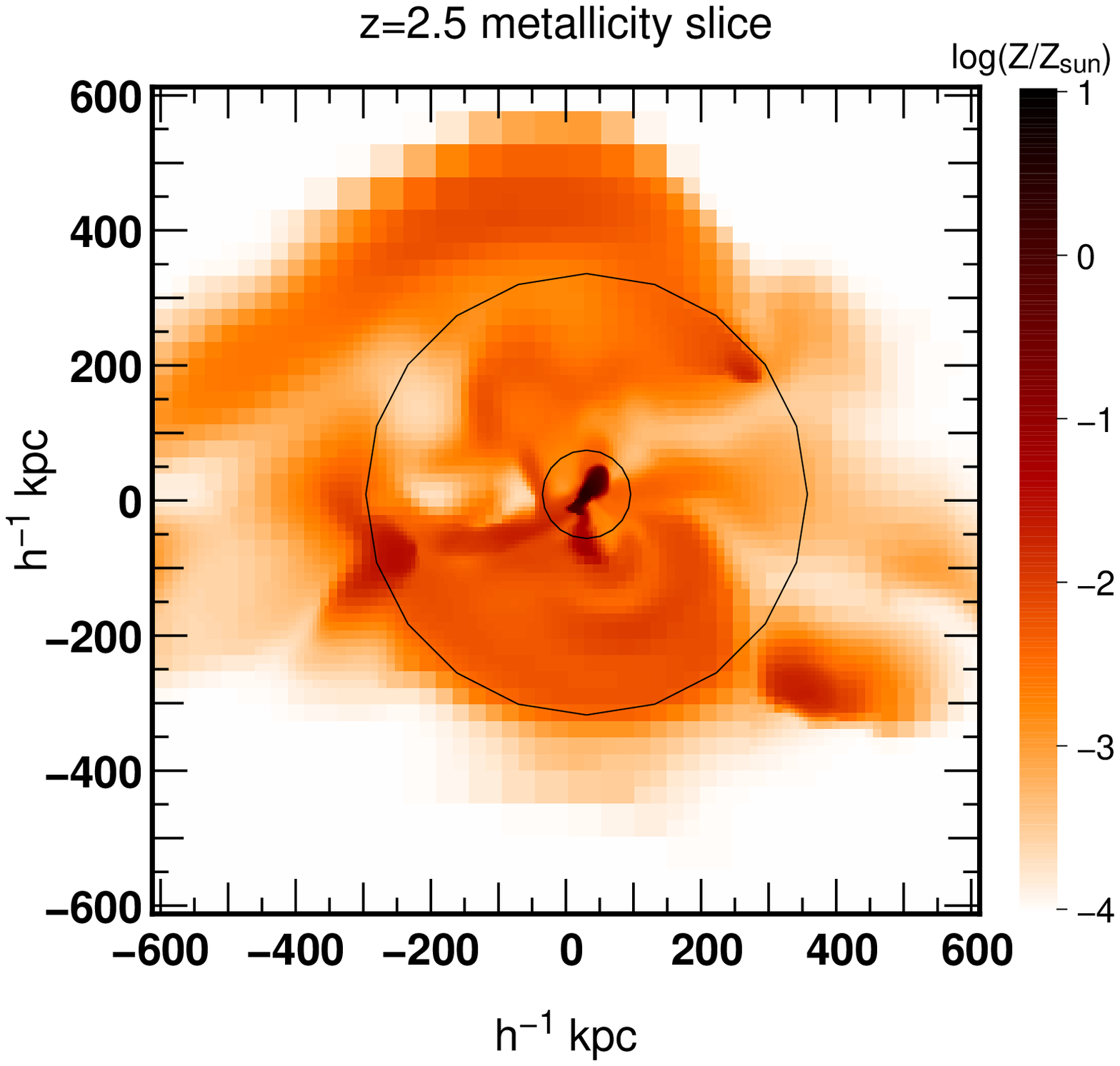}}
\end{tabular}
\caption{Maps of the physical properties of the gas for a typical $M=2 \times 10^{12}
\Msun$ halo. {\em Left:} $z=4$,  {\em right:} $z=2.5$. {\em Top:} Projected density, {\em middle:} temperature slice, {\em bottom:} metallicity slice. The large (small) black circle shows the location of $\Rv$ ($0.2 \Rv$) for the central galaxy. Note the disappearance 
of  filaments at  low redshift.}
\label{f:space}
\end{figure*}

This critical  mass seems  to increase sharply  with redshift (radially  averaged case, top  panel of Fig. \ref{f:hotcold}). Note
that at  redshift 5.4,  it can only  be guessed  since no halo  in the
simulation is massive enough to have $\langle \fhot \rangle \ge 0.5$.
This evolution  is the  signature of  a  gradual
disappearance of cold radially extended features, like filaments, in the
massive haloes between $z=5.4$ and  $z=2$.  This is illustrated by \Fig{space}, showing  maps  of a  typical  halo of  mass
$\MDM=2\times 10^{12}\Msun$  at $z=4$  (left) and another  halo of
the same mass at redshift  $z=$2 (right). While the former features clear
filaments streaming into the inner halo, the latter lies at the centre
of a hot  bubble, with no apparent filaments  inside the Virial radius
(large black  circle). The critical  mass defined  by the
accretion transition in the entire halo  marks the disappearance of  cold streams.
It is therefore called here $\Mst$.

On the  contrary, the critical mass at the inner halo ($0.2 \Rv$), i.e. the mass of halos with
$\fhot \ge 0.5$  shows only  a slow
variation with redshift,  if any.  It indicates that  accretion in the
inner parts of the halo switches to  the hot mode as soon as $\MDM \ge
10^{11.5-12} \Msun$ at all redshifts considered,  while the  outer part of  the halo can  still be
dominated  by the  cold  mode.   Again, this  is  well illustrated  by
\Fig{space}. The inner part of the halo (inner circle) is shock
heated at  both redshifts,  although the radius  of the accretion  shock is
much larger in  the low redhift snapshot. At  high redshift, the accretion
shock coexists  with cold streams coming  from the outer  parts of the
halo. The  critical mass  defined  at the inner halo marks  the  appearance of  an accretion  shock  around  the
galaxy. It is therefore called here $\Msh$.

%************************ REFORMATTED:2columns by 3 lines ********************
An important  parameter in our  approach is the density  threshold we
used to  remove clumpy satellites  from the analysis.  We  checked that
our conclusions in terms  of transition masses and average metallicity
of  the hot  phase are  robust to changes of this parameter  by  repeating our
measurements with a  lower density threshold. The main  effect of this
extra gas removal is to  reduce the fraction of high metallicity, cold
gas  in the  vicinity  of galaxy  satellites,  without any  noticeable
effect on the metallicity of the hot phase and on the critical masses.

%\section{Discussion}
\section{Comparison to earlier theoretical modelling}
\label{s:DB06}
The physics  of accretion has been investigated by several authors in the past. It is
insightful to review their results in light of our measurements. Fig. \ref{f:literature}  shows the evolution of $\Mst$  with
redshift. For $z=2$  it is approximately $10^{11.5} \Msun$  and it increases
sharply with increasing $z$.  At $z \ge 4$ and  above, even the most
massive  haloes   in  the  simulation  are  still   dominated  by  cold
accretion. However, a rough extrapolation  of the $z=4$ curves of Fig. \ref{f:hotcold} yields a
transition  mass of  about $\approx  10^{12.7}  \Msun$. Qualitatively,
this  behaviour is  in agreement  with  the evolution  of $\Mst$  with
redshift  as derived by DB06. According  to their study,
filaments exist only in $\rm{M} \le \Mst$ haloes. However, their figure
7 shows  that $\MstDB(z=2)=10^{12.4} \, \Msun$,  which is significantly
larger  than  what we  find.  We will  now  show  that correcting  the
metallicity assumptions of DB06 with our measurements (low metallicity in the filaments) can  reconcile  these discrepant  values.

\subsection{Approximation of DB06 model}

In order to compare our results with DB06 calculations, we will first build a crude interpolation of their $\Mst$ and $\Msh$ for any redshift, radius, and metallicity. Analysing  the
dependence  of  $\Msh$  with  respect  to the  problem  parameters  in
equation (34)  of DB06, we  can isolate the dependence  in metallicity
as:
\begin{equation}
\log(\Msh(r,Z_0,z))=0.7 \log(Z_0)+ A(r,z) \, ,
\label{eq:Mshmodel}
\end{equation}
where $Z_0=Z/ \Zsun$ at $z=0$,  and $A(r,z)$, which contains the dependence of $\Msh$ with respect to radius and redshift, can be tabulated from fig. 2 and 4 of DB06.
Then, taking the log  of their equation (40), we see  that $\Mst$ is defined
with respect to the critical mass for shock stability, $\Msh$, as
\begin{equation}
\log(\Mst) = 2 \log(\Msh)- \log(\Mps) - \log(3) \, ,
\end{equation}
where $\Mps$ is the typical dark  matter halo mass at a given redshift
as computed using the formalisms of \citet{lahav91,carroll92,mo02} and
following Appendix 3 of DB06. Since $Z_0$ is required at $z=0$, we need to extrapolate our measurements at $z=2$ to $z=0$. To do so, we will for consistency use the same  enrichment rate s=0.17 as in DB06, as derived from \citet{delucia04}. 

\subsection{Hot shocks}
Since the volume filling factor of the cold phase is in general negligible compared to that of the hot phase when looking at the halo as a whole, only the hot phase metallicity $\Zhot$ is relevant when evaluating the stability of the hot shock.
At the inner halo, Fig. \ref{f:Zhot} or the fit of Eq. \ref{eq:Zrfit} show that $\Zhot(0.1\Rv,z=2)\approx -1$. This translates into $\log(\Zhot(0.1\Rv,z=0)/\Zsun)=-0.65$, which is rather close to the metallicity assumption of DB06, i.e. $\log(Z_0/\Zsun)=-1$ at 0.1 $\Rv$. As a consequence, we expect the critical hot shock mass $\Msh$
computed by DB06 to match our measurements well, provided that we compute
and measure $\Msh$ at the same depth in the halo, i.e. at 0.2 $\Rv$. We take as an average metallicity at this radius $\log(\Zhot(0.2\Rv,z=2)/\Zsun) = -1.7$, which is intermediate between the 3 mass bins shown in Fig. \ref{f:Zhot}.
This translates into $\log(\Zhot(0.2\Rv,z=0)/\Zsun) \approx -1.4$. The shock mass $\Msh$ obtained with this metallicity and radius is shown in Fig. \ref{f:literature}. A good
agreement is indeed achieved between this model and our measurements. We also agree with
a quasi constant $\Msh$ as found by \cite{keres05} and \cite{BD07} in SPH simulations,
although the absolute normalization seems to differ. However, their
methodology is quite different (they analyse the temperature history of their
gas particles) and their metallicity is uniform and constant throughout the whole simulation.

\subsection{Cold streams}
\begin{figure}
\includegraphics[width=1.\linewidth,clip]{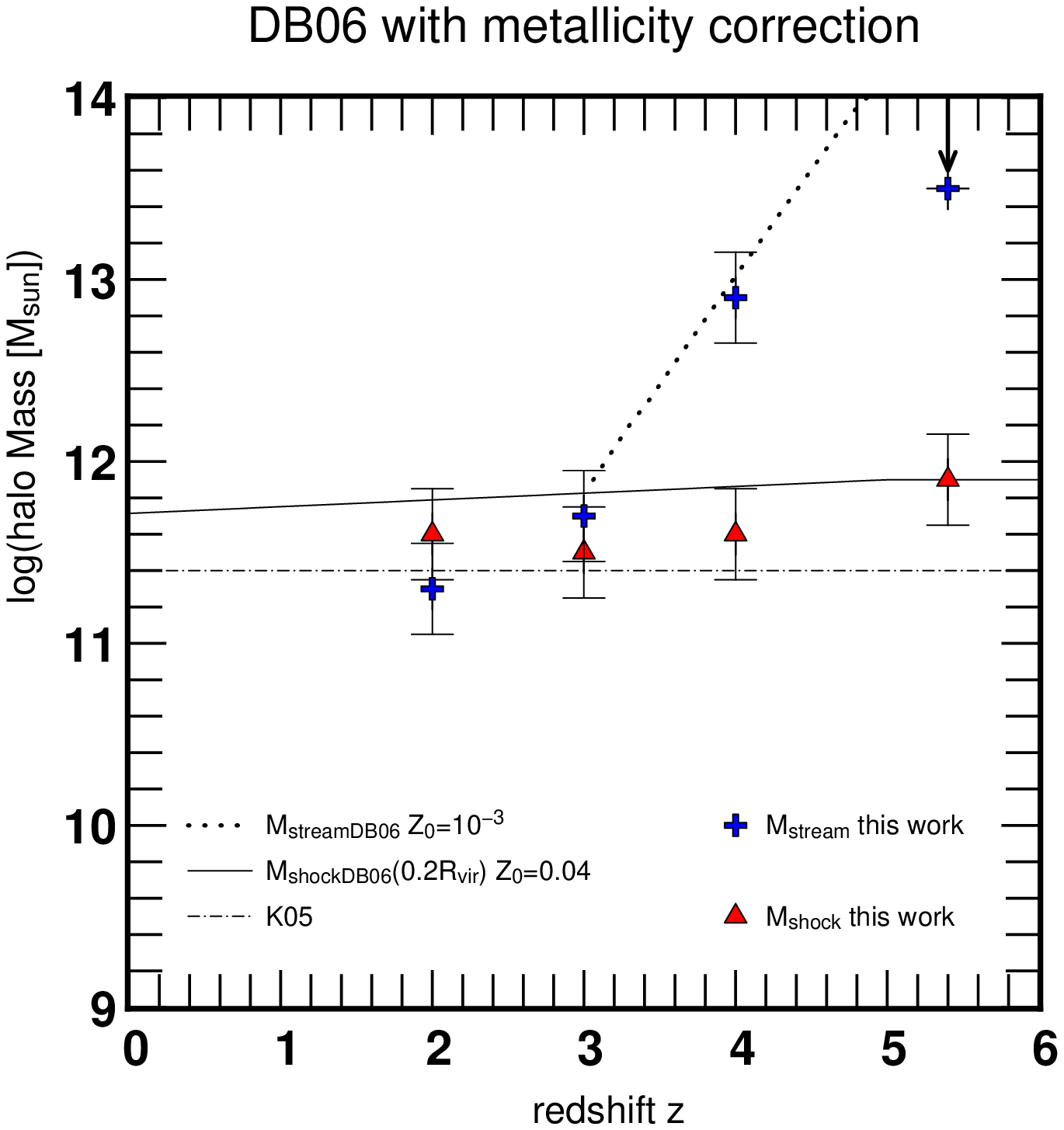}
\caption{
Evolution of $\Msh$ and $\Mst$ with redshift, from our measurements and
comparison to analytical modelling. The solid line shows DB06 prediction for
$\Msh (0.2\Rv)$ assuming $Z_0/\Zsun=0.04$, while the dotted line shows their
prediction for $\Mst$ with a modified metallicity
$Z_0/\Zsun=10^{-3}$. Finally, the dash dotted line shows the constant
transition mass reported by Kere{\v s} et al. (2005). The error bars on our measurements are equal to the widths of
the mass bins used. The arrow pointing to the $\Mst (z=5.4)$ indicates that
this point is given as a lower limit only, as can be estimated from  Fig. 4.
}
\label{f:literature}
%done with Mshock5.i on tikka
\end{figure}

%In DB06, $\Msh$ is calculated at the inner halo (0.1 $\Rv$), and consequently $\Mst$ is linked to this small region as well. However, it is clear  from Fig. \ref{f:space} that for steady
%cold streams to be  stable in a hot gas bubble they  need to be stable
%all  the  way  up  to the outer halo. 
In DB06, $\Mst$ is related to $\Msh$, which strongly  depends  on
metallicity via  the cooling function, as  also shown by  fig. 10 of
BD03  and  fig.  2  of  DB06. As established from \Sec{bimod} and \Fig{bigDF4}, the metallicity of cold streams is rather low. At $z=4$, for example, the metallicity queue of the cold accretion mode extends down to $Z = 10^{-8} \, \Zsun$ and possibly lower, while the cold high-$Z$ accretion consists of the surroundings satellite galaxies gas disks, rather than genuine cold streams. The efficiency of radiative cooling decreases towards low metallicity, and fig. 13 of \cite{SD93} shows that at $Z/\Zsun=10^{-3}$ the cooling properties of the gas are already those of a primordial mixture. We use this value of the metallicity for the DB06 modelling of $\Mst$. Doing so shifts the disappearance of cold streams to earlier times (i.e. higher $z$) with respect to the original assumption of DB06 ($Z_0=0.1\Zsun$), as a consequence of less effective radiative cooling.
\Fig{literature} shows that this modelling of the low-$Z$ cold streams then match our measurements rather well.
%On the one hand, using our value  $Z_0=10^{-3.5}$, we get $\MstDB$
%shown  in \Fig{literature}, which agrees well with our measurements. On the other hand, the latter rule out the $Z_0=0.1$ hypothesis at $\Rv$. This
%difference can  be attributed to a  lower efficiency  of the
%radiative  cooling with  decreasing metallicity. 
The value plotted for $\Mst$ at
$z=5.4$ is an extrapolation and is given only as a lower limit (indicated by the
arrow).

\subsection{Discussion}
We see that while the relatively high metallicity of DB06 at the inner
halo is a fair assumption, a hundredfold lower metallicity must nonetheless be
assumed for the gas filaments, which then have radiative properties similar to a primordial mixture.
Once the importance and the effect of these assumptions has been accounted
for, the consistency between the MareNostrum measurements and the theory of DB06
is remarkably good, and shows that their analytical approach indeed seems to capture
the essence of our (nonetheless limited) understanding of the processes involved in gas accretion physics as modelled by the simulation. 

The agreement on $\Mst$ also suggests that the main process driving the stability of cold filaments in a hot halo is the competition between a compressive increase of temperature in the filament on the one hand, 
and radiative cooling on the other hand. 
This is an important point since several other hydrodynamical processes are expected to be at work, such as Kelvin-Helmholtz instabilites, which are not modelled in DB06. However, the MareNostrum simulation is not tailored to resolve them. \cite{klein94} showed that at least 100 cells per cloud  radius are needed to resolve interface instabilities involved in cloud destruction by shock waves. However, it has been argued that the gas filaments can be stabilized by the underlying dark matter stream, and could totally prevent Kelvin-Helmholtz instability from appearing. Filaments can also be stabilized if the flow is supersonic, which is indeed the case: for the halo shown in Fig. \ref{f:space}, the inflowing filament from the lower right quadrant of the central galaxy has a Mach number of 2-3. Moreover, the density contrast of the cold gas stream and the average density at the Virial radius  is also an important parameter. DB06 note that it is not a well-constrained quantity at the moment, although measurements are under way, and could well shift within a whole decade around the current assumptions in their modelling. Decreasing this ratio could have a similar effect to decreasing the metallicity, in making the cold filaments disappear earlier (i.e. denser filaments are more stable).

Finally, we recall that the low $Z$ measured in the filaments could be a result of the limited resolution of the simulation. Indeed, additional fragmentation down to the lowest masses allowed by the UV radiation field could still  take place in the filaments, forming high-redshift dwarf galaxies, and the resulting star formation episodes could enrich the filaments.

%Although this is a different context, this criterion is illustrative of the cell size needed to resolve Kelvin-Helmholtz instability in cold gas filaments. Here the finest cells are about one kpc large. But this refinement level is attained in galaxy discs only, and not in filaments. It has been argued that the gas filaments can be stabilized by the underlying dark matter stream. In that case the Kelvin-Helmholtz instability could disappear totally. In this respect, DB06 notes that $\Mst$
%However, no study of this problem has been conducted and thus it is still an open question. In this respect, although our results agree with DB06, both approaches might still miss essential physics involved in the stability of filaments.

%Finally, in DB06, the metallicity of the cold streams is linked to the
%metallicity of the gas as a whole, and thus partly inherits its cooling properties. While this simplification makes the problem
%analytically more tractable (it allows the authors to circumvent
%mixing issues, together with the fact that the mass rate in the cold and hot phase may differ
%significantly), it is in disagreement with the results shown in Fig. 
%\ref{f:bigDF4}. Allowing the two gas phases to have different
%metallicities could lead to a significant improvement of the DB06 model.

\section{Comparison to observations}

It  is  difficult  to  find   observables  to  compare  our
measurements to since our measurements ``only'' reach down to $z=2$ while  observational studies of the
galaxy  bimodality  are generally  restricted to  $z\le1.5$, and the bimodality becomes clear only at later times. Also, the definition of the
critical masses constrained by theoreticians and observers have little in
common: theoreticians have access to quantities that are in general impossible to measure with existing or even future 
observing facilities (for instance the mass accretion rate of cold
gas at $\Rv$), while observers have access to enormous volumes of the universe and events which would be impossible to simulate with the sufficiently high spatial and temporal resolution required for imposing useful constraints. Hence, shortcuts are taken. They
always involve external assumptions, and the relevance of the quantities
being compared must be carefully examined. Here  we discuss  our
results in the  light of the observed transition  and quenching masses
measured  by \cite{bundy06}, hereinafter B06.\\
 The stellar to dark matter mass ratio $\Mstellar  / \MDM$ used  to estimate  the observed stellar  mass $\Mstellar$ from the dark matter halo mass $\MDM$ and its evolution with redshift is subject  to  large  uncertainties and is a hotly debated topic. It is linked to the evolution of the Tully-Fisher relation \citep{tully-fisher,bell01}. However we recall that the latter rather focuses on the $\Mstellar/{\rm M_{total}}$ ratio where the total mass ${\rm M_{total}}=\Mstellar + {\rm M_{gas}} + \MDM$ is the sum of the stellar, gas and dark matter masses. However, the sum is always strongly dominated by the dark matter mass for the systems and radii we consider here. It  seems reasonable  to  expect  that the ratio $\Mstellar / \MDM$ increases  with decreasing redshift \citep{rettura06,kannappan07},  as more gas is turned  into stars. However, a non-evolving  $\Mstellar / \MDM$ could also be in agreement with both observations \citep{bamford06,boehm07,atkinson07} and numerical studies \citep{portinari07}. In our simulation a ratio $\Mstellar / \MDM \approx 50$ seems reasonable at  $z=2$ for structures in the mass range considered here. 
%{\pier{ check references in more detail}}
\\
%\subsection{The ${M_tr-z}$  relation}
B06 measured the evolution with redshift of the transitional mass $\Mtr$ marking the transition from the blue to red sequence of galaxies. This study is based on a sample of 8000  DEEP2 galaxies between $0.4 \le z \le 1.4$. The galaxies are divided into two groups (star forming and passive), ideally according to their SFR, with a limiting SFR=$0.2 \Msun/{\rm yr}$. However,
the [OII] emission line generally used to derive the SFR \citep{kewley04} falls in the
DEEP2 survey wavelength range only for $z\ge 0.75$. As a
consequence, the colour index (U-B) is also used as a proxy for the SFR as measured
from [OII], and an absolute magnitude-dependent colour cut is made as an alternative to the SFR cut \citep{dokkum00}. The authors show that indeed, very few galaxies with (U-B)$\ge
0.2$ are forming stars. However, there is a large fraction ($\approx 30 \%$) of galaxies
with (U-B)$\le 0.2$ which are passive. This means that the sample
of passive galaxies will suffer only limited pollution from star-forming
galaxies, while the star-forming sample contains a significant
fraction of passive galaxies ($\approx 30 \%$ at $z=[0.75-1]$ as can be seen from their
Fig. 1). Once the red and blue galaxy groups are defined (through the colour or SFR criterion), the authors compute the mass functions of the blue and red groups and the contribution of blue and red galaxies to the total mass function. They show that the massive end of the mass function is dominated by red galaxies while the fainter end is dominated by blue galaxies. They then define two transition masses:
\begin{enumerate}
\item{$\Mtr$ is the mass for which the mass function of red galaxies equals that of the blue galaxies, i.e. both groups contribute equally to the total mass function. $\Mtr$ shows only a moderate dependence on the red/blue separation criterion. It is found to increase from $10^{10.5} \Msun$ to $10^{10.8} \Msun$ between $0.4 \le z \le 1.4$.}
\item{$\Mqb$ is the mass for which the fraction of blue galaxies drops below $1/3$ (which means the contribution of blue galaxies is half the contribution of the red galaxies to the total mass function). The authors claim it represents the mass at which star formation ``quenches''. It increases from $10^{10.73} \Msun$ to $10^{11.23} \Msun$ between $0.4 \le z \le 1.4$.}
\end{enumerate}
Various other investigations consider different photometric filters, colour cuts, SFR estimates or mass estimates, but rely on similar methodology \citep{arnouts07,hopkins07}. Given the complexity of the method and the number of steps and assumptions involved, it is clear that the physical meaning of the observed transition masses can be quite far  from that of the transition masses we compute from our gas accretion measurements. However, it is admitted that they correspond to a real change in the evolution of SFR or specific SFR with galaxy mass. The details are difficult to assess. For instance, does the transition mass signify a total shutdown of star formation or a smooth decline? Such details are bound to be fuzzy because we are looking at populations whose properties have an intrinsic dispersion, as they have different histories, environment etc. Having recalled these difficulties, we now proceed to make a number of remarks:
\begin{enumerate}
\item{Using a stellar to dark matter mass ratio of $\Mstellar/\MDM=50$ we get ${\rm M^{\bigstar}_{\rm shock}}\approx 10^{9.9} \Msun$ at $z=2$, which is significantly smaller than $\Mtr$ found in B06 at any epoch. Moreover, since $\Mtr$ increases with $z$, the disagreement is bound to be stronger if we were to observe $\Mtr(z=2)$. A similar disagreement is seen for $\Mst$.}
\item{On the other hand, we can define a mass $\Mq$ where the cold gas accretion rate on the central galaxy drops to zero, also corresponding to the mass where $\fcold=0$ at the inner halo, instead of $\fcold=0.5$ as for $\Msh$. We find $\Mq=10^{12.8}\Msun$, which translates to $\approx 10^{11.1}\Msun$ in stellar mass. Although $\Mqb$ is expected to increase between $z=1.4$ and $z=2$, the agreement is remarkable. It is also roughly consistent with the results of \cite{pozzetti03,fontana04}, based on the K20 survey. This raises the question as to what we should adopt as threshold in $\fcold$. Indeed we see that moving this threshold can result in a shift of more than a decade in the masses obtained. There is actually little reason for expecting $\fcold=0.5$ haloes to match observed transition masses precisely.}
% other than sheer naivety.}
\item{We find a difference between $\Msh$ and $\Mq$ of about $1.2-1.3$ dex. This is actually a measure of the sharpness of the transition from cold to hot accreting haloes. This contrasts with the finding of B06 where $\Mtr-\Mqb \approx 0.2-0.4$ dex, which suggests a much sharper transition. In future modelling works and observational studies, computing the sharpness in mass of the transition from blue to red galaxies could help constraining the mechanism responsible for the transition.}
\item{We find a quasi-constant $\Msh$, with only very slight evolution with $z$, compatible with the findings of earlier theoretical works, both numerical and analytical. In contrast, both $\Mtr$ and $\Mqb$ evolve strongly with $z$. On the other hand, it is difficult to check the evolution of $\Mq$ with redshift, since the only epoch where $\fcold \approx 0$ haloes exist in the simulation is $z=2$. A naive linear extrapolation of the evolution of $\fcold$ at larger redshifts suggests a quasi-constant $\Mq$, but this is simply impossible to check and would require simulating a larger box in order to get haloes massive enough to reach $\fcold=0$ earlier in the life of the universe.}
\item{If we believe that $\Mq$ has a physical meaning similar to $\Mqb$ this apparent difference in their evolution with redshift is problematic. Moreover, a stellar to dark matter ratio $\Mstellar / \MDM$ evolving with $z$ would make this issue even worse. Indeed, a constant $\Mq$ combined with a decreasing $\Mstellar /\MDM$ with $z$ would lead to a {\em decreasing} $\Mq(z)$, which would then be in even stronger disagreement with observations.}
\item{On the other hand, $\Mst$ does evolve with $z$, with a slope similar to that of $\Mtr$ and $\Mqb$. It would be interesting to compute the mass at which the cold fraction $\fcold$ drops to zero at the outer halo because such a mass may be comparable in absolute normalisation {\em and} in slope to $\Mqb$. However, this is difficult to even estimate from the current simulation, and the corresponding masses are clearly out of reach as can be seen on Fig. \ref{f:hotcold}.}
\end{enumerate}
Finally, it becomes clear that in order to compare our theoretical transition masses to observed ones, one should use the same transition definitions in the simulation as in the observations. This involves building catalogues of galaxies (rather than dark matter haloes), computing their luminosities and colours according to the age and metallicity distributions of their stars using stellar population models such as \cite{bc03,PEG,PEG-HR,coelho07}. The resulting colour-magnitude diagrams of the galaxy population could then be compared directly to observations such as the SDSS $M_r / (u-r)$ distribution \citep{baldry04}. This is the approach adopted in SAMs such as those of \cite{cattaneo06}.
The same colour cuts as in the observations \citep{arnouts07,dokkum00} could then be applied and the corresponding red and blue galaxy mass functions constructed, along with the corresponding transition masses. 
 Since the colour bimodality of galaxies falls into place only after $z \le 1.5$ (the last redshift bin of \cite{arnouts07} shows no bimodality in (NUV-r')$/$K at all), it should not be expected that the bimodality can be clearly seen in a colour-magnitude diagram even at the lowest redshift of the simulation. Moreover, since the AGN feedback now often proposed as the origin of the galaxy bimodality has not been implemented in the MareNostrum simulation, it might never appear even if we could continue the simulation down to lower redshifts. In any case, the above colour-magnitude diagram synthesis and the corresponding colour cuts have to be carried out quantitatively to check for the presence/absence of a galaxy bimodality in the MareNostrum simulation.

\section{Conclusions}
%\begin{enumerate}
We used the Horizon-MareNostrum galaxy formation simulation to study the processes involved in gas accretion on galaxies. We introduced
 mass accretion rate weighted statistics that allow us to quantify the mass accretion rates as a function of gas temperature and metallicity.

Gas accretion is bimodal both in temperature and in metallicity, defining a hot and a cold accretion mode. The cold accretion mode is associated with a combination of metal poor filamentary accretion and dense metal-rich satellite galaxy disc surroundings, while the hot accretion mode features strong chemical heterogeneity, and a radius-dependent metallicity.

%The cooling properties of the hot gas are affected by this inhomogeneity and deviate strongly from those of a gas having the average metallicity of the hot accreted phase. 
We give an analytical fit to the metallicity of the hot accretion mode as a function of radius, which will be relevant for future SAMs.

We define $\Msh$ and $\Mst$, the halo masses for which cold and hot accretion contribute equally, at the inner halo and within the whole halo, respectively. Haloes more massive than $\Msh$ develop stable hot shocks, but may still possess cold gas filaments nourishing the galaxy disc. For halo masses larger than $\Mst$, these filaments disappear. $\Msh$ is found to be quasi-constant with $z$, while $\Mst$ increases sharply. Our results for $\Msh$ are in agreement with the analytical stability calculations of DB06, and our metallicity measurements support their original assumption {\em at the inner halo}. Conversely, our determinations of $\Mst$ disagree with their original predictions. We show that assuming a low metallicity for the filaments, as we measured ($10^{-3} \, \Zsun$ is similar to primordial abundances as far as cooling is concerned) brings their model in agreement with our results.

This suggests that, as long as the stability of hot shocks and cold streams is assumed to be mainly driven by a  competition between compression and radiative cooling, their analytical modelling is accurate within this model;  it then depends critically on assumptions made about the metallicity, via the cooling function. 
%In this respect the fit to the metallicity of the hot phase given here could be an important input to such analytical models.

We propose that in addition to the transition masses, two other observables should always be considered by future SAMs or numerical investigations of the origin of the galaxy bimodality: 
\begin{enumerate}
\item{the sharpness of the transition $\log(\Mq/\Msh)$, whose observable counterpart could for instance be the log ratio of the quenching mass to the transition mass as defined in B06.}
\item{The evolution of transition/quenching masses with redshift.}
\end{enumerate} 
Modelling these two quantities should  be helpful in pinpointing the nature of star formation quenching in galaxies.

Comparing the transition masses we obtain to observed transition masses is a difficult task, and we found only marginal agreement. The diffuse cold gas supply drops to zero at the inner halo for an estimated stellar mass ${\rm M}^{\bigstar}_{\rm quench}\approx 10^{11.1} \Msun$ at $z=2$, which is remarkably close to the quenching masses observed by B06. Unfortunately, we are not able to constrain the evolution of $\Mq$. However, we note that the evolution of the observed quenching mass is similar to the evolution of $\Mst$.
In this respect, the agreement between measured and observed quenching masses suggests that simulating realistic galaxy populations does not necessarily require more ingredients than the physics already modelled in the MareNostrum simulations. We recall here that no AGN feedback has been taken into account in this work.
To be more conclusive, better constraints on $\Mq$ and its evolution are needed. These could be obtained by means of zoom-simulations of smaller boxes centred on massive haloes to lower $z$. Since $\Mq$ is determined mostly by the few most massive haloes of the simulation, a larger box or more realizations of a box of the same size are required to improve the statistics.

Conversely, additional ingredients may become necessary in order to prevent clumpy gas accretion from reaching the galaxy centre when it is significant (a study of the contribution of clumpy gas accretion will be carried out in a forthcoming paper). But they need not be in the form of AGN feedback. Although stripping of the hot halo of infalling satellite galaxies is properly resolved in the MareNostrum simulation, the interface instabilities cold disc/hot gas and cold filaments/hot gas are not. These could be the missing physics.

Similarly,  some energy input into the hot component  might be required to maintain its temperature and  avoid cooling flows at later stages. \cite{DB08} proposed that ``gravitational quenching'' could be a solution. It involves keeping the inner halo gas hot through interactions with cold dense gas clumps (drag), allowing to transfer the potential energy of these infalling clouds to the inner halo.
%\item{fcold =0 at inner est une conditino necessaire pour quencher alors que fcold = 0 at outer halo est une conditino suffisante.}
%\item{Par contre on aura peut etre besoin d'un nouvel ingredient physique pour eliminer l'accretion clumpy, mais it does not need be AGN feedback: It could be that unresolved Kelvin helmmholtz instabilities acting on accreted satellites do a good job at removin gtheir fresh gas. This would also have an effect on the cold streams, which would then likely be disrupted earlier than we predict here.}
A crucial step arising from our study is that of the interplay
between the  hot gas  bubble and cold  streams/clouds. Are hot gas bubbles
able  to  disrupt  filaments  connected  to the  disc  via  electronic
conduction, turbulence, Kelvin-Helmhotz instability? Are cold streams
immune to disruption thanks to increased pressure (feeling pressure of
the hot gas) and thus higher density and consequently higher radiative
cooling  efficiency  or  the  underlying  dark  matter  stream?  Can
electronic conduction be impeded  by local magnetic fields of suitable
intensity? Is  there a  pressure/temperature/halo mass  for  which a
given cold stream of  given density/velocity/DM flux will be disrupted
by these instabilities? It will be decades before cosmological simulations with the same size as the MareNostrum simulation will have the resolution required to resolve interface instabilities. Hence, as a first step, more restricted, idealised experiments are needed in order to investigate these phenomena.

\section*{Acknowledgements}
We thank the referee Y. Birnboim for his useful comments which helped us improve the paper.
The authors thankfully acknowledge the computer resources, technical expertise and assistance provided
by the Barcelona Supercomputing Centre - Centro Nacional de Supercomputacion.
This work  was  performed  within  the framework of  the Horizon
collaboration  ({\tt http://www.projet-horizon.fr}). We thank D.~Aubert,  A. ~Dekel, P.~Clark, S.~Colombi, J.~Devriendt, J.~Forero-Romero, S.~Glover, N.~Maddox and D. Leborgne for useful comments and helpful suggestions. 
We would  also like to thank D.~Munro for
freely distributing his Yorick programming language (available at
\texttt{http://yorick.sourceforge.net/}) PO was    supported by a grant from the Centre National de la Recherche Scientifique (CNRS) and a grant from the Deutsches Zentrum f\"ur Luft und Raumfahrt (DLR).

\bibliographystyle{mn2e}
\bibliography{ocvirk_rev1}
%%%%%%%%%%%%%%%%%%%%%%%%%%%%%%%%%%%%%%%%%%%%%%%%%%%%%%%%%%%%%%%%%%%%%%%%%%%%%%%
%

\label{lastpage}
\end{document}